# A Fundamental Scale of Descriptions for Analyzing Information Content of Communication Systems

**Gerardo Febres * and Klaus Jaffe**

Laboratorio de Evolución, Universidad Simón Bolívar, Sartenejas, Baruta, Miranda, 1080, Venezuela; E-Mail: kjaffe@usb.ve

* Author to whom correspondence should be addressed; E-Mail: mail@gfebres.com; Tel.: +58-414-333-7513; Fax: +58-212-730-2782.

---

**Abstract:** The complexity of the description of a system is a function of the entropy of its symbolic description. Prior to computing the entropy of the system's description, an observation scale has to be assumed. In texts written in artificial and natural languages, typical scales are binary, characters, and words. However, considering languages as structures built around certain preconceived set of symbols, like words or characters, limits the level of complexity that can be revealed analytically. This study introduces the notion of the fundamental description scale to analyze the essence of the structure of a language. The concept of Fundamental Scale is tested for English and musical instrument digital interface (MIDI) music texts using an algorithm developed to split a text in a collection of sets of symbols that minimizes the observed entropy of the system. This Fundamental Scale reflects more details of the complexity of the language than using bits, characters or words. Results show that this Fundamental Scale allows to compare completely different languages, such as English and MIDI coded music regarding its structural entropy. This comparative power facilitates the study of the complexity of the structure of different communication systems.

---

## 1. Introduction

The understanding of systems and their complexity requires accounting for their entropy. The emergence of information upon the scale of observation has become a topic of discussion since it reveals much of the systems' nature and structure. Bar Yam [1] and Bar-Yam *et al.* [2] have proposed the concept of complexity profile as a useful tool to study systems at different scales. Among others, Lopez-Ruiz *et al.* [3], and Prokopenko [4] focus on the change of the balance between the system disorder and self-organization for different scales of observation. In a different approach, Gell-Mann [5] considers complexity as a property associated to the irregularities of the physical system. But Gell-Mann sees both randomness and order as manifestations of regularity, and therefore quantities that offer the possibility for reducing the length of a description and hence the computed complexity of a system.

These complexity concepts are all evaluated using arbitrarily selected symbol scales. The selected observation scale depends on the communication system used in the description; for example, systems described with human natural languages are prone to be analyzed with the characters and words scales because they hold the most meaning for humans. When the analysis of information is in the context of its transmission, it is common to find binary codes as the base of study. A possible consequence of this preselected scale of observation is the possible inclusion of our assumptions about the system's structure, which skews our interpretation about system properties.

Many studies have evaluated the entropy of descriptions based on a preconceived scale; in 1997 Kontoyiannis [6] evaluated the description entropies at the scale of characters; in 2002 Montemurro and Zanette [7] studied the entropy as a function of the word-role; more recently Savoy [8], and Febres *et al.* [9,10] have studied the impact of the style of writing over entropy speeches using the word as the unit of the scale. In 2009 Piasecki and Plastino [10] studied entropy as a function of a 2-dimensional domain. They explored the effects of multivariate distributions and calculate the entropy associated to several 2D patterns. All these studies share the same direction; assume a space for a domain and a scale and compute the entropy. The strategy of the present study is to set the same problem in a reversed fashion: given an entropy descriptor of a multivariate distribution defined for some domain space, what would be the best way to segment that domain space in order to reproduce the known entropy descriptor? The answer to this question would have a twofold value: (a) an indication to the scale that best represents the system expression as the distribution of sizes of the space segments, and (b) an approximation to the algorithmic complexity of the description.

Algorithmic complexity as a concept does not consider the observation scale [5,11]. Algorithmic complexity—also called Kolmogorov's complexity—is the length of the shortest string that completely describes a system. Since the shortest string is a characteristic impossible to guarantee, algorithmic complexity has been regarded as an unreachable figure. Nevertheless, estimating complexity by searching for a nearly uncompressible description of a system, would have the advantage of being independent of the observation scale. In fact, a method to search for a nearly uncompressible description could be achieved by adjusting the observation scale until the process discovers the scale that best comprises the original description. The result would lead to an approximation to the algorithmic complexity of the system.

While these previous studies assume symbols as characters or words, in our present study we leave freedom to group adjacent characters, to form symbols in order to comply with a higher hierarchy criterion, as is the minimization of the entropy. This study develops a series of algorithms to recognize the set of symbols that, according to their frequency, leads to a minimum entropy description. The method developed in this study mimics a simplified communication system's evolution process. The proposed algorithm is tested with short example of English text, and two descriptions, the first is an English text and the second, a sound musical instrument digital interface (MIDI) file. This representation of the components may convey a description of a system and its structural essence.

## 2. A Quantitative Description of a Communication System

A version of Shannon's entropy formula, generalized for communication systems comprised of $D$ symbols, is used to compute quantity of information in a descriptive text. To determine the symbols that make up the sequential text, a group of algorithms were developed. These algorithms are capable of recognizing the set of symbols which form the language used in the textual description. The number of symbols $D$ represents not only the diversity of the language but also the fundamental scale used for the system description.

### *2.1. Quantity of Information for a $D'$nary Communication System*

We refer to language as the set of symbols used to construct a written message. The number of different symbols in a language will be referred as the diversity $D$.

To compute the entropy $h$ of a language, that is, the entropy of the set of $D$ different symbols, used with a probability $p_i$ to form a written message, we use the Shannon's entropy expression, normalized to produce values between zero and one:

$$h = -\sum_{i=1}^{D} p_i \cdot log_D \, p_i \,, \tag{1}$$

Note that the base of the logarithm is equal to the language's diversity $D$, whereas classical Shannon's expression uses 2 as the base of the logarithm; also equal to the diversity of the binary language that he studied. Researchers such as Zipf [12], Kirby [13], Kontoyiannis [6], Gelbukh and Sidorov [14], Montemurro and Zanette [7], Savoy [15], Febres, Jaffe and Gershenson [9] and Febres and Jaffe [16], among others, have studied the relationship between the structure of some human and artificial languages, and the symbol probability distribution corresponding to written expressions of each type of language.

All these studies assume symbols as characters or words, in our present study we leave freedom to group adjacent characters, to form symbols in order to comply with the minimization of the entropy $h$ as expressed in Equation (1). In the following sections we explain this optimization problem, and our approach to find a solution reasonably close to the set of symbols that produce an absolute minimum entropy.

### *2.2. Scale and Resolution*

We propose a quantitative concept of scale: the scale of a system equals the diversity of the language used for its description. Thus, for example, if a picture is made with all available colors in an 8-bit-color map of pixels, then the diversity of the color language of the picture would equal $2^8$, and the scale of the picture description, considering each color as a symbol, would be also $2^8$. Another example would be a binary language, a scale 2 communication system made up of only two symbols. Notice we have used the term "communication system" to refer to the media used to code information.

Interestingly, the system's description scale is determined, in first place, by the observer, and in a much smaller degree by the system itself. The presumably high complexity of a system, functioning with the actions and reactions of a large number of tiny pieces, simply dissipates if (a) the observer or the describer fails to see the details, (b) the observer or describer is not interested the details, and prefers to focus on the macroscopic interactions that regulate the whole system's behavior, or (c) the system does not have sufficient different components, which play the role of symbols here, to refer to each type of piece. It is clear that any observed system scale implies the use of a certain number of symbols. It is also clear that the number of different symbols used in a description is linked with our intuitive idea of *scale*. There being no other known quantitative meaning of the word *scale,* we suggest its use as a descriptor of languages by specifying the number of symbols forming them.

Resolution specifies the maximum accuracy of observation and defines the smallest observable piece of information. In the computer coded files we used to interpret descriptions, we consider the character as the smallest observable and non-divisible piece of information.

Let $E$ denote the physical space that a symbol or a character occupies, and let the sub-index signal the object being referred to. Thus, considering a written message $\boldsymbol{M}$, constructed using $D_{\boldsymbol{M}}$ different symbols $Y$ as $\boldsymbol{M} = \{Y_1, Y_2, \dots, Y_{D_{\boldsymbol{M}}}\}$, we would say the message $\boldsymbol{M}$ occupies the space $E_{\boldsymbol{M}}$ and each symbol $Y_i$ occupies the space $E_{Y_i}$. We define the length of all characters equal one. Therefore $E_{C_i} \equiv 1$ for any $i$. Finally, if the number of characters in a message is $N$, each symbol $Y_i$ appears $F_{Y_i}$ times within the message, and the symbol diversity is $D_{\boldsymbol{M}}$, we can write the following constraints over the number of characters, symbols and the space they occupy:

$$E_{\boldsymbol{M}} = \sum_{i=1}^{D_{\boldsymbol{M}}} F_{Y_i} \cdot E_{Y_i} = \sum_{i=1}^{N} E_{C_i} = N \,. \tag{2}$$

### *2.3. Looking for a Proper Language Scale*

We see the scale of a language as the set of finite symbols that "best" serves to represent a written message. The qualification "best" refers to the capacity of the set of symbols to convey the message with precision in the most effective way.

Take for example the western natural languages. Among their alphabets, there are only minor differences; too few differences to explain how far from each other those languages are. As Newman [17] observes, some letters may be the basic units of a language, but there are other units formed by groups of letters.

Chomsky's syntactic structures [18], later called context-free grammar (CFG) [19] offers another representation of natural language structure. The CFG describes rules for the proper connections among words according to their specific function within the text. Thus, CFG is a grammar generator useful to study the structure of sentences. Chomsky himself treats a language as an infinite or finite set of sentences. CFG works at a much larger scale than the one we are looking for in this study.

Regarding natural languages it is common to think that a word is the group of characters within a leading and a trailing blank-space. At some time a meaning was assigned to that word, and thereafter the word's meaning, as well as its writing, evolves and adopts a shape that works fine for us, the users of that language. Zipf's principle of least effort [14] and Flesch's reading ease score [20] certainly give indications about the mechanisms guiding words, as written symbols, to reduce the number of characters needed to be represented.

From a quantitative linguistics perspective, this widely accepted method for recognizing words offers limited applicability. Punctuation signs, for example, have a very precise meaning and use. The frequency of their appearance in any western natural language compete with the most common words in English and Spanish [21]. However, punctuation signs are very seldom preceded by a blank-space and are normally written with just a single character, which promotes the false idea that they function like letters from the alphabet; they do not. They have meaning as well as common words have. Another situation revealing the inconvenience of this natural but too rigid conception of words, is the English contraction when using the apostrophe. It is difficult to count the number of words in the expression "they're". How many words are there, one or two? See Febres *et al.* [21] for a detailed explanation on English and Spanish word recognition and treatment for quantification purposes.

Intuitively the symbols forming a description written using some language, should be those driving the whole message to low entropy when computed as the function of the symbols frequency. In this situation the message is fixed as fixed is also the text and the quantity of information it conveys. Then, there appears to be a conflict: while the information is constant because the message is invariant, any change to the set of symbols considered as basic units, alters the computed message entropy, as if the information had changed; it has not. To solve this paradox, we return to the question asked at the beginning of this section about the meaning of "best" in the context of this discussion. From the point of view of the message emitter, the term "best" considers the efficiency to transmit an idea. This is what Shannon's work was intended for: to determine the amount of information, estimated as entropy, needed to transmit an idea. From the reader's point of view the economy of the problem works different. The reader's problem is to interpret the message received to maximize the information extracted. In other words, the reader focuses on the symbols which turn the script as an organized, and therefore easier to interpret message. If the reader is a human and there are words in the message, the focused symbols are most likely words because those are the symbols that add meaning for this kind of reader. But if there existed the possibility to select another set of symbols which makes the message look even more organized, the reader would rather use this set of symbols because it would require less effort to read.

In conclusion, what the reader considers "best" is the set of symbols that maximizes the organization of the message while for the sender the "best" means the set of symbols needed to minimize the disorder of the message and thus the quantity of information processed. These statements are expressed as objective functions in Equations (3) where the best set of symbols is named $\boldsymbol{B}$, the message is $\boldsymbol{M}$, the message entopy is $h_{\boldsymbol{M}}$ and the message organization is $(1 - h_{\boldsymbol{M}})$:

$$\begin{aligned} \text{Senders}'\text{objective: } & \min_{\boldsymbol{B}} h_{\boldsymbol{M}} \\ \text{Receiver}'\text{s objective: } & \max_{\boldsymbol{B}} \ (1 - h_M) = \min_{\boldsymbol{B}} h_{\boldsymbol{M}} \end{aligned} \tag{3}$$

Following this reasoning, "best" means the same for both sides of the communication process. This may have important implications when considering languages as living organisms or colonies of organisms. Both parts of the communication process push the language to evolve in the same direction:

augmenting self-organization and the reducing of entropy of the messages. Both come together. Self-organization can be seen as one of the evolving directions of languages. Thus, self-organization is an indirect way to measure how deeply evolved a language is and what its capacity is to convey complex ideas or sensations. Finally, an objective function to search the most effective set of symbols—the set with minimal entropy—to describe a language has been found. It will be used to recognize the set of symbols that best describes a language used to write a description.

### *2.4. Language Recognition*

Consider a description consisting of a message $\boldsymbol{M}$ built up with a sequence of $N$ characters or elementary symbols. The message $\boldsymbol{M}$ can be treated as an ordered set of characters $C_i$ as:

$$\boldsymbol{M} = \{C_1, C_2, \dots, C_N\}. \qquad (4)$$

No restriction is imposed over the possibility of repeating characters. Consider also the language $\boldsymbol{B}$, consisting of a set of $D_{\boldsymbol{B}}$ different symbols $Y_i$, each formed with a sequence of $E_{Y_i}$ consecutive characters found with probability $P(Y_i) > 0$ in message $\boldsymbol{M}$. Thus:

$$\boldsymbol{B} = \{Y_1, Y_2, \dots, Y_{D_B}, P(Y_i)\}. \qquad (5)$$

$$Y_i = C_j, C_{j+1}, \dots, C_{j+E_{Y_i}-1}\}, 1 \le i \le D_{\boldsymbol{B}}, 1 \le j \le N - E_{Y_i} + 1. \qquad (6)$$

The symbol probability distribution $P(Y_i)$ can be obtained dividing the frequency distribution $f_i$ by the total number of symbols $N$ in the message:

$$P(Y_i) = \frac{f_i}{N} \qquad (7)$$

Language $\boldsymbol{B}$, used to convey the message $\boldsymbol{M}$, can now be specified as the set of $D_{\boldsymbol{B}}$ different symbols and the probability density function $P(Y_i)$ which establishes the relative frequencies of appearance of the symbols $Y_i$. Each symbol $Y_i$ is constructed with a sequence of contiguous characters as indicated in Equation (6). The set of symbols that describes the message $\boldsymbol{M}$ with the least entropy comes after the solution of the following optimization problem:

$$\min_{\boldsymbol{B}} \; -\sum_{i=1}^{D_{\boldsymbol{B}}} \frac{F_{Y_i} \cdot E_{Y_i}}{N} \cdot log_{D_{\boldsymbol{B}}} \frac{F_{Y_i} \cdot E_{Y_i}}{N},$$

Subject to: $\boldsymbol{B} = \{Y_1, Y_2, \dots Y_i \dots, Y_{D_B}, \boldsymbol{P}(Y_i)\}$, for $i = 1, 2, \dots, D_{\boldsymbol{B}}$

$$Y_i = \left\{C_j, C_{j+1}, \dots, C_{j+E_{Y_i}-1}\right\}, \text{for } i = 1, 2, \dots, D_{\boldsymbol{B}} \; and \; j = 1, 2, \dots N - E_{Y_i} + 1, \qquad (8)$$

$$\textstyle\sum_{i=1}^{D_{\boldsymbol{B}}} F_{Y_i} \cdot E_{Y_i} = N,$$

$$F_{Y_i} \ge 1, \; E_{Y_i} \ge 1, for \; i = 1, 2, 3, \dots, D_{\boldsymbol{B}}.$$

The resulting language will be the best in the sense that it is the set of symbols that offers a maximum organization of the message. The symbol lengths will range from a minimum to a maximum defining a distribution of symbol lengths characteristic of this scale of observation which is referred to as the *Fundamental Scale*.

## 3. The Algorithm

The optimization problem (8) is highly nonlinear and restrictions are coupled. A strategy for finding a solution has been devised. It is a computerized process compound of text-strings processing, entropy calculations, text-symbol ordering and genetic algorithms. Given a description consisting of a text of $N$ characters, the purpose of the algorithm is to build a set of symbols $\boldsymbol{B}$ whose entropy is close to a minimum. The process forms symbols by joining as many as $V$ adjacent characters in the text. A loop where $V$ is kept constant, controls the size of the symbols being incorporated to language $\boldsymbol{B}$. The process ends when the maximum symbol length of $V_{mx}$ characters is considered to form symbols. We add a sub-index to language $\boldsymbol{B}_V$ to indicate the symbol size $V$ considered at each stage of its construction. We have defined several sections of the algorithm and we named them according to their similarity with a system where each symbol appears and ends up being part of a language, only if it survives the competence it must stand against other symbols. A pseudo-code of the fundamental scale algorithm is included in Appendix A.

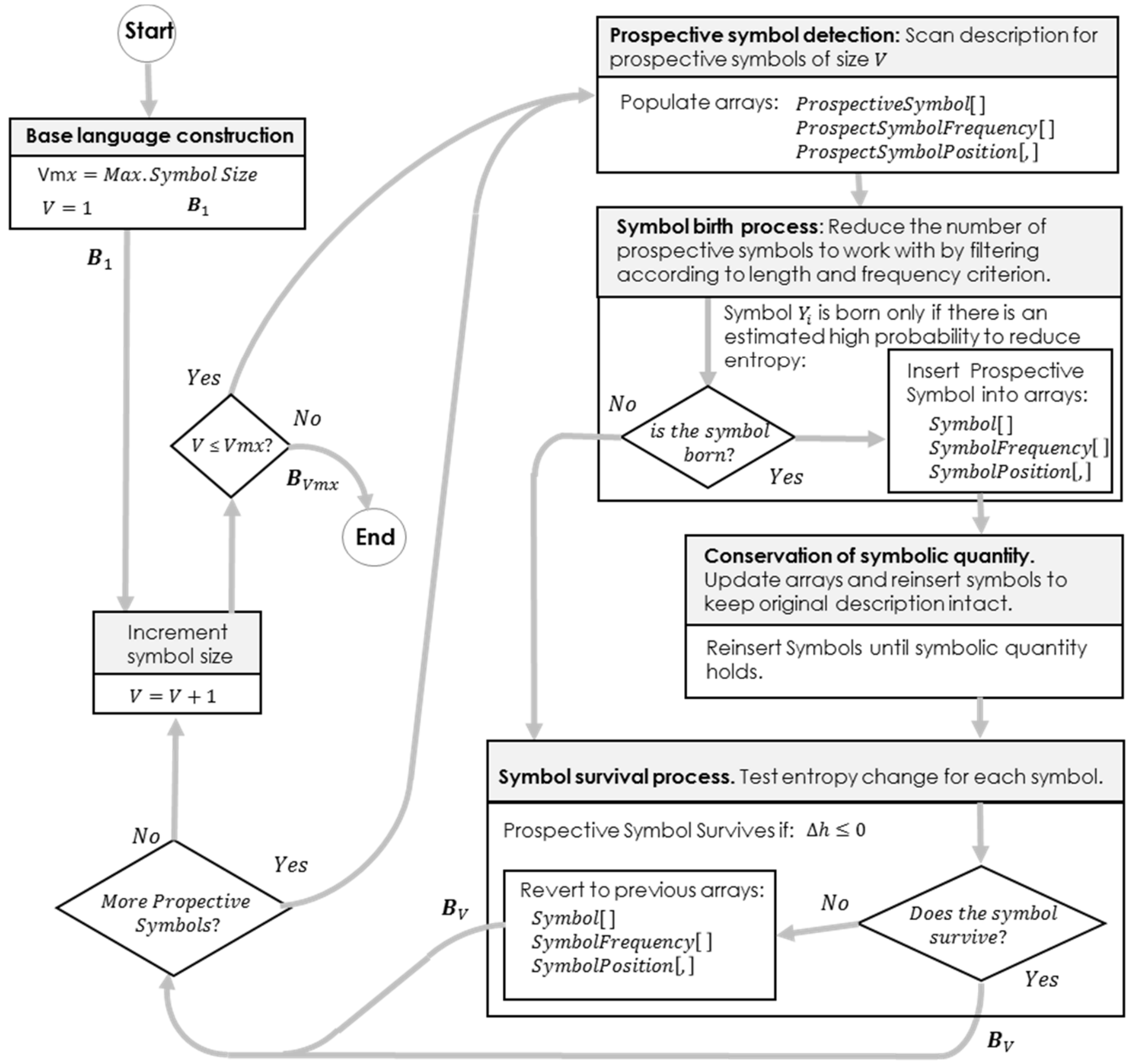

**Figure 1.** Major components of the fundamental scale algorithm.

*3.1. Base Language Construction*

In the first stage, the message $\boldsymbol{M}$ is separated into single characters. The resulting set of characters along with their frequency distribution constitute the first attempt to obtain a good language and it will be denoted as $\boldsymbol{B}_1$. The sub-index indicates the maximum length that any symbol can achieve.

*3.2. Prospective Symbol Detection*

The prospective symbol detection consists of scanning the text looking for strings of exactly $V$ characters. All $V$-long strings are considered as prospective symbols to join the previously constructed language $\boldsymbol{B}_{V-1}$ made of strings of up to $V-1$ characters. The idea is to find all possible different $V$-long strings present in the message $\boldsymbol{M}$, which after complying with some entropy reduction criteria, would complement language $\boldsymbol{B}_{V-1}$ to form language $\boldsymbol{B}_V$.

To cover all possibilities of character sequences forming symbols of length equal to $V$, several passes are done over the text. The difference from one pass to another is the character where the initial symbol starts, which will be called the *phase* of the pass. Figure 2 illustrates how the strategy covers all possibilities of symbol instances for any symbol size specification $V$.

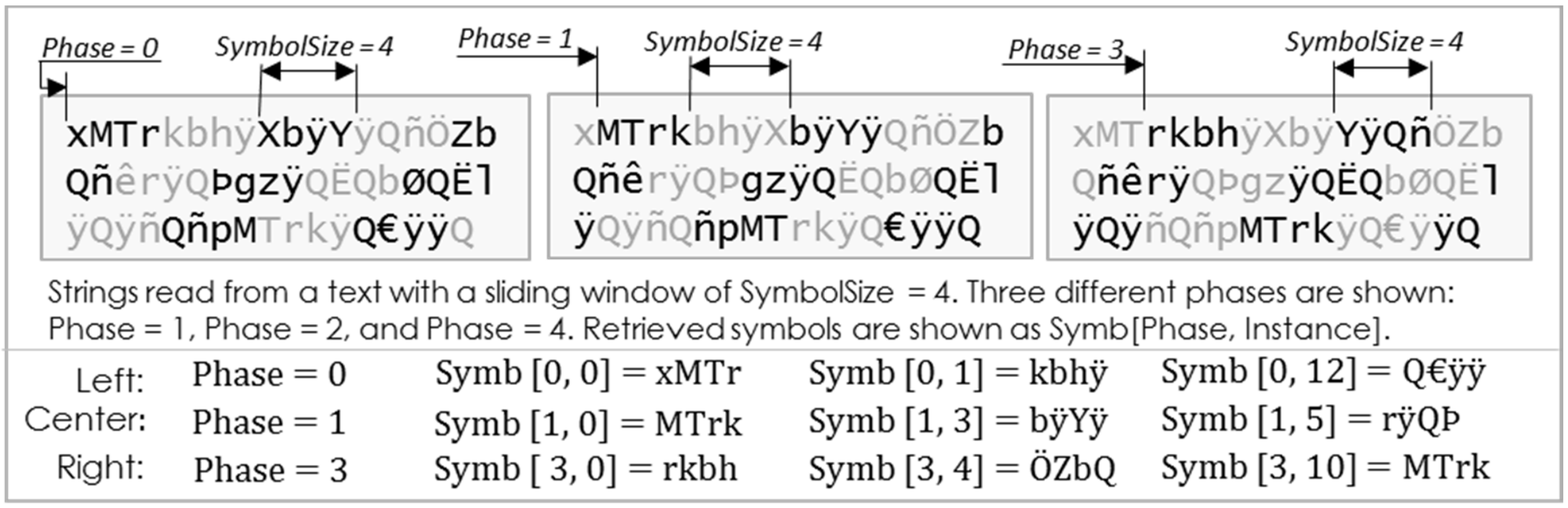

**Figure 2.** Examples of reading a text to recognize prospective symbols with a sliding window of *SymbolSize* = 4 and reading *Phase* = 0, 1, and 3. *Phase* = 2 not shown. The message: "*xMTrkbhÿXbÿYÿQñÖZbQñêrÿQÞgzÿQËQbØQËlÿQÿñQñpMTrkÿQ€ÿÿQ*".

*3.3. Symbol Birth Process*

Prospective Symbols detected in the previous stage whose likelihood to be an entropy reducer symbol is presumed too low, are discarded and never inserted as part of the language. Interpreting entropy Equation (1) as the summation of contributions of the uncertainty due to each symbol, we can intuit that minimum total uncertainty—minimum entropy—occurs when each symbol uncertainty contribution is about the same. Thus, any Prospective Symbol must be close to the average uncertainty per symbol in order to have some opportunity to actually reduce the entropy after its insertion. The average contribution of the uncertainty $u_i$ for symbol $i$ can be estimated as:

$$u_i = -p_i \, log_{D_{\boldsymbol{B}_V}} \, p_i = \frac{h}{D_{\boldsymbol{B}_V}}, \tag{9}$$

This leads us to look for symbols complying with condition shown in Equation (10), and save processing time whenever a prospective symbol is not within a 2λ-width band of around the average uncertainty value:

$$\frac{h}{D_{\boldsymbol{B}_V}} - \lambda < u_i < \frac{h}{D_{\boldsymbol{B}_V}} + \lambda . \qquad (10)$$

Parameter $\lambda$ can be adjusted to avoid improperly rejecting entropy reducer symbols or to operate in the safe side at the expense on processing time.

### *3.4. Conservation of Symbolic Quantity*

The inclusion of prospective symbols into the arrays of symbols representing the language $\boldsymbol{B}$, is performed to avoid the overlap of the newly inserted symbols and the previous language existing symbols. Therefore, every time a prospective symbol is inserted into the stack of symbols, the instances of former symbols occupying the space the new symbols, must be released. Sometimes this freed string is only a fraction of a previously existing symbol. Thus, the insertion of a symbol may produce a break up of other symbols, generating empty spaces for which recovered symbols must be reinserted in order to keep the original text intact.

### *3.5. Symbol Survival Process*

A final calculation is performed to confirm the entropy reduction achieved after the insertion of a symbol into the language being formed. Those symbols not producing an entropy reduction, are rejected and the Language $\boldsymbol{B}$ is reverted to its condition prior to the last insertion of a symbol.

### *3.6. Controlling Computational Complexity*

The computational complexity of this algorithm is far beyond polynomial. A rough estimation sets the number of steps required above the factorial of the diversity of the language treated. Thus, segmenting the message into shorter pieces, allows the algorithm to find a feasible solution and to keep affordable processing times for large texts. This strategy is in fact a sort of parallel processing which significantly reduces the algorithm's computational complexity down to becoming an applicable tool. A complex system software platform has been developed along with this study to deal with the complexities of this algorithm, and the structure needed to maintain record of every symbol of each description within a core of very many texts. This experimental software, is named *Monet* and a brief description of it can be found in [21].

The noise introduced when cutting the original description in pieces, is limited. At most two symbols may be fractured for each segment. Very low compared to the number of symbols making each segment. The algorithm calculates the entropy of each description chunk. But, as Grabchak *et al*. [22] explain, the estimation of the description's entropy must consider the bias introduced when short text samples are evaluated. Taking advantage of the extensive list of symbols and frequencies available and organized by means of the software *Monet*, we used the alternative of calculating the description entropy using the joint sets of symbols for each description partition, an then forming the whole description. As a result, no bias has to be corrected.

## 4. Tests and Results

In order to compare the differences obtained when observing a written message at the scales of characters, words and the fundamental scale, we designed an Example Text. Table 1 shows the symbols obtained after the analysis of the Example Text at the three observation scales used in this study. The entropies calculated at the scales of characters and words were 0.81 and 0.90 respectively, the entropy at the fundamental scale was 0.76; an important reduction of the information required to describe the same message.

These results also get along with our intuition. Clearly, the selection of a certain character-string as a fundamental symbol, is favored by the frequency of appearance of the string of characters. As a result, the "*space character*" (represented as ø in the table) is recognized as the most frequent fundamental symbol. It indeed is an important structural piece in any English text, since it defines the beginning and the end of natural words. The length of the string of characters also favors the survival of the symbol in its competence with other prospective symbols. The string "*describ*", for example, appears twice in the Example Text and the algorithm recognized it as a symbol. On the other hand, the 11-char long string "*An adverb*" also appears two times, but the algorithm found it more effective in reducing the overall entropy, to break that phrase apart and increase the appearances of other symbols. A similar case is that of the word "*adverb*", which appears in nine instances (not including those written with the first capital letter) on the Example Text. But the entropy minimization problem found a more important entropy reduction by splitting the word "*adverb*" in shorter and more frequent symbols as "*dv*" (10 times), or the characters as "*e*" (70 times), "*a*" (40 times), ), "*r*" (33 times), and "*b*"(12 times).

In another experiment, we contrasted two different types of communication systems by performing tests over full real messages. The first test is based on a text description written in English and the second in test based on the text file associated to music coded using the MIDI format. The English text is a speech by Bertrand Russell given in 1950 during the Nobel Prize ceremony. The MIDI music is a version of the 4th movement of Beethoven's ninth symphony. The sizes of these descriptions are near the limit of applicability of the algorithm. English descriptions of 1300 words or less can be processed in short times of less than a minute. Larger English texts have to be segmented using the control computational complexity criteria mentioned in Section 3.6 to reach reasonable working times. Bertrand Russell's speech was fractioned in seven pieces. For MIDI music files, the processing times show an attitude of sharp increase starting for music pieces lasting about 3 min. The version of 4th movement of Beethoven's ninth symphony used, is a 25 min long piece. It was necessary to process it by fractioning in 20 segments.

To reveal the differences of descriptions when observed at different scales, symbol frequency distributions were produced. For the English text, characters, words and the fundamental scale were applied. For the MIDI music text distributions at character and fundamental scale were constructed. Words do not exist as scale for music. The corresponding detailed set of fundamental symbols can be seen in Appendix B. The frequency distributions were ordered upon the frequency rank of the symbols, thus the obtained were Zipf's profiles.

**Table 1.** Results of the analysis of the Example Text at the three scales studied.

Example Text: symbol sets at different scales.

-What is an adverb? An adverb is a word or set of words that modifies verbs, adjectives, or other adverbs. An adverb answers how, when, where, or to what extent, how often or how much (e.g., daily, completely). Rule 1. Many adverbs end with the letters "ly", but many do not. An adverb is a word that changes or simplifies the meaning of a verb, adjective, other adverb, clause, or sentence expressing manner, place, time, or degree. Adverbs typically answer questions such as how?, in what why?, when?, where?, and to what extent?. Adverbs should never be confused with verbs. While verbs are used to describe actions, adverbs are used describe the way verbs are executed. Some adverbs can also modify adjectives as well as other adverbs.

$F_Y$ = Frequency, $E_Y$ = Space occupied, $N$ = Message length, ø =space, $V_{mx}$ = max. symb. length = 13

| Char scale | | | Word scale | | | | | | Fundamental scale | | | | | | | |
|---|---|---|---|---|---|---|---|---|---|---|---|---|---|---|---|---|
| *D = 38 h = 0.8080* | | | Diversity *D = 82* Entropy *h = 0.9033* | | | | | | Diversity *D = 80 h = 0.7628* | | | | | | | |
| *d = 0.0486 N = 782* | | | Specific diversity *d = 82* Length *N = 171* | | | | | | Specific diversity *d = 1384* Length *N = 578* | | | | | | | |
| **Idx.** | **Symbol** | ***FY*** | **Idx.** | **Symbol** | ***FY*** | **Idx.** | **Symbol** | ***FY*** | **Idx.** | **Symbol** | ***FY*** | ***EY*** | **Idx.** | **Symbol** | ***FY*** | ***EY*** |
| 1 | ø | 169 | 1 | , | 21 | 41 | completely | 1 | 1 | ø | 100 | 1 | 41 | ul | 2 | 2 |
| 2 | e | 86 | 2 | . | 11 | 42 | ) | 1 | 2 | e | 70 | 1 | 42 | wi | 2 | 2 |
| 3 | a | 45 | 3 | or | 7 | 43 | Rule | 1 | 3 | a | 40 | 1 | 43 | io | 2 | 2 |
| 4 | s | 44 | 4 | adverbs | 7 | 44 | 1 | 1 | 4 | s | 36 | 1 | 44 | ie | 2 | 2 |
| 5 | r | 44 | 5 | ? | 6 | 45 | end | 1 | 5 | t | 36 | 1 | 45 | im | 2 | 2 |
| 6 | t | 39 | 6 | adverb | 5 | 46 | letters | 1 | 6 | r | 33 | 1 | 46 | whe | 2 | 3 |
| 7 | o | 34 | 7 | verbs | 4 | 47 | ly | 1 | 7 | o | 22 | 1 | 47 | øan | 2 | 3 |
| 8 | d | 32 | 8 | how | 4 | 48 | but | 1 | 8 | n | 21 | 1 | 48 | dif | 2 | 3 |
| 9 | n | 30 | 9 | an | 4 | 49 | do | 1 | 9 | , | 18 | 1 | 49 | uch | 2 | 3 |
| 10 | h | 28 | 10 | what | 4 | 50 | not | 1 | 10 | h | 17 | 1 | 50 | ,øc | 2 | 3 |
| 11 | i | 25 | 11 | is | 3 | 51 | changes | 1 | 11 | b | 12 | 1 | 51 | anyø | 2 | 4 |
| 12 | v | 21 | 12 | a | 3 | 52 | simplifies | 1 | 12 | dv | 10 | 2 | 52 | wordø | 2 | 5 |
| 13 | b | 21 | 13 | other | 3 | 53 | meaning | 1 | 13 | d | 9 | 1 | 53 | describ | 2 | 7 |
| 14 | w | 21 | 14 | to | 3 | 54 | verb | 1 | 14 | c | 8 | 1 | 54 | .øAdverb | 2 | 8 |
| 15 | , | 21 | 15 | the | 3 | 55 | adjective | 1 | 15 | u | 7 | 1 | 55 | ød | 1 | 2 |
| 16 | c | 17 | 16 | as | 3 | 56 | clause | 1 | 16 | l | 6 | 1 | 56 | øv | 1 | 2 |
| 17 | l | 16 | 17 | are | 3 | 57 | sentence | 1 | 17 | ? | 6 | 1 | 57 | word | 1 | 4 |
| 18 | . | 11 | 18 | word | 2 | 58 | expressing | 1 | 18 | wh | 6 | 2 | 58 | yø | 1 | 2 |
| 19 | u | 11 | 19 | of | 2 | 59 | manner | 1 | 19 | w | 5 | 1 | 59 | ma | 1 | 2 |
| 20 | m | 10 | 20 | that | 2 | 60 | place | 1 | 20 | i | 5 | 1 | 60 | f | 1 | 1 |
| 21 | y | 10 | 21 | adjectives | 2 | 61 | time | 1 | 21 | . | 4 | 2 | 61 | ns | 1 | 2 |
| 22 | f | 7 | 22 | when | 2 | 62 | degree | 1 | 22 | g | 4 | 1 | 62 | An | 1 | 2 |
| 23 | ? | 6 | 23 | where | 2 | 63 | typically | 1 | 23 | x | 4 | 1 | 63 | w | 1 | 2 |
| 24 | A | 5 | 24 | extent | 2 | 64 | answer | 1 | 24 | ly | 4 | 2 | 64 | b, | 1 | 2 |
| 25 | g | 5 | 25 | with | 2 | 65 | questions | 1 | 25 | m | 4 | 1 | 65 | v | 1 | 1 |
| 26 | p | 5 | 26 | “ | 2 | 66 | such | 1 | 26 | verbs | 4 | 5 | 66 | - | 1 | 1 |
| 27 | x | 4 | 27 | used | 2 | 67 | in | 1 | 27 | y | 3 | 1 | 67 | ( | 1 | 1 |
| 28 | j | 3 | 28 | describe | 2 | 68 | why | 1 | 28 | p | 3 | 1 | 68 | ) | 1 | 1 |
| 29 | W | 2 | 29 | many | 2 | 69 | and | 1 | 29 | dj | 3 | 2 | 69 | R | 1 | 1 |
| 30 | " | 2 | 30 | - | 1 | 70 | should | 1 | 30 | øof | 3 | 3 | 70 | 1 | 1 | 1 |
| 31 | - | 1 | 31 | set | 1 | 71 | never | 1 | 31 | ctiv | 3 | 4 | 71 | M | 1 | 1 |
| 32 | ( | 1 | 32 | words | 1 | 72 | be | 1 | 32 | .øA | 2 | 3 | 72 | q | 1 | 1 |
| 33 | ) | 1 | 33 | modifies | 1 | 73 | confused | 1 | 33 | . | 2 | 1 | 73 | S | 1 | 1 |
| 34 | R | 1 | 34 | answers | 1 | 74 | While | 1 | 34 | W | 2 | 1 | 74 | ho | 1 | 2 |
| 35 | 1 | 1 | 35 | often | 1 | 75 | actions | 1 | 35 | " | 2 | 1 | 75 | øm | 1 | 2 |
| 36 | M | 1 | 36 | much | 1 | 76 | way | 1 | 36 | ow | 2 | 2 | 76 | ng | 1 | 2 |
| 37 | q | 1 | 37 | € | 1 | 77 | executed | 1 | 37 | me | 2 | 2 | 77 | if | 1 | 2 |
| 38 | S | 1 | 38 | e | 1 | 78 | Some | 1 | 38 | le | 2 | 2 | 78 | in | 1 | 2 |
| | | | 39 | g | 1 | 79 | can | 1 | 39 | øi | 2 | 2 | 79 | on | 1 | 2 |
| | | | 40 | daily | 1 | 80 | also | 1 | 40 | pl | 2 | 2 | 80 | si | 1 | 2 |
| | | | | | | 81 | modify | 1 | | | | | | | | |
| | | | | | | 82 | well | 1 | | | | | | | | |

Table 2 shows the length $N$, the diversity $D$ and the entropy $h$ obtained for these two descriptions analyzed at several scales and Figure 3 shows the corresponding Zipf's profiles for Bertrand Russell's speech English speech and Beethoven's 9th Symphony's 4th movement. Both descriptions' profiles are presented at the scales they were analyzed: character-scale and the fundamental scale for both, English and music, and the word-scale only for English.

In Figures 3a and 3b, the character scale exhibit the smallest diversity range. Taking only the characters as allowable symbols, leaves out any possibility of combination to form more elaborated symbols and excluding any possibility of representing how the describing information of a system arranges to create what could be loosely called the "language genotype". Allowing the composition of symbols as the conjunction of several successive characters, dramatically increases the diversity of symbols.

The selection of the symbols to build an observation scale holding the criteria of minimizing the resulting frequency distribution entropy, bounds the final symbolic diversity in a scale while capturing a variety of symbols that represents the way characters are organized to represent the language structure. The fundamental scale appears as the most effective scale, since with it, the original message can be represented with the most compressed information, expressed as the lowest entropy measured for all scales in both communication systems evaluated.

**Table 2.** Details of two descriptions used to test the fundamental scale method.

| | | **Name of scale** | | | | | | | | |
|---|---|---|---|---|---|---|---|---|---|---|
| | | **Characters** | | | **Fundamental** | | | **Words** | | |
| **Text tag** | **Communication System** | **Length *N*** | **Diversity *D*** | **Entropy *h*** | **Length *N*** | **Diversity *D*** | **Entropy *h*** | **Length *N*** | **Diversity *D*** | **Entropy *h*** |
| .Bertrand Russell 1950.NobelLecture | English | 32,621 | 68 | 0.7051 | 26,080 | 1227 | 0.5178 | 6476 | 1590 | 0.8215 |
| Beethoven. Symphony9.Mov4 | MIDI Music | 103,564 | 160 | 0.6464 | 84,645 | 2824 | 0.4658 | not defined | | |

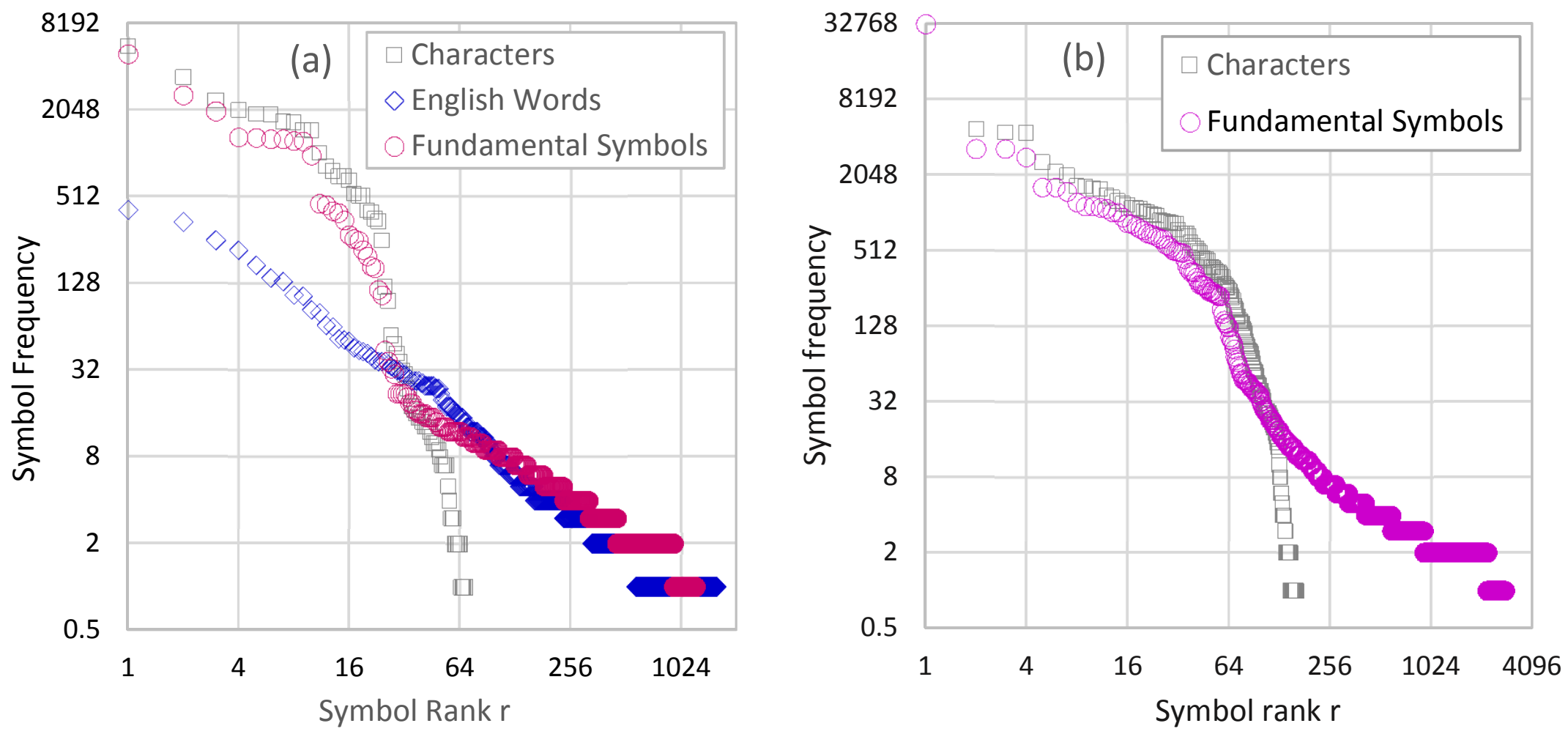

**Figure 3.** Symbol profiles for an English text (**a**) and a MIDI music text (**b**) at different scales of observation.

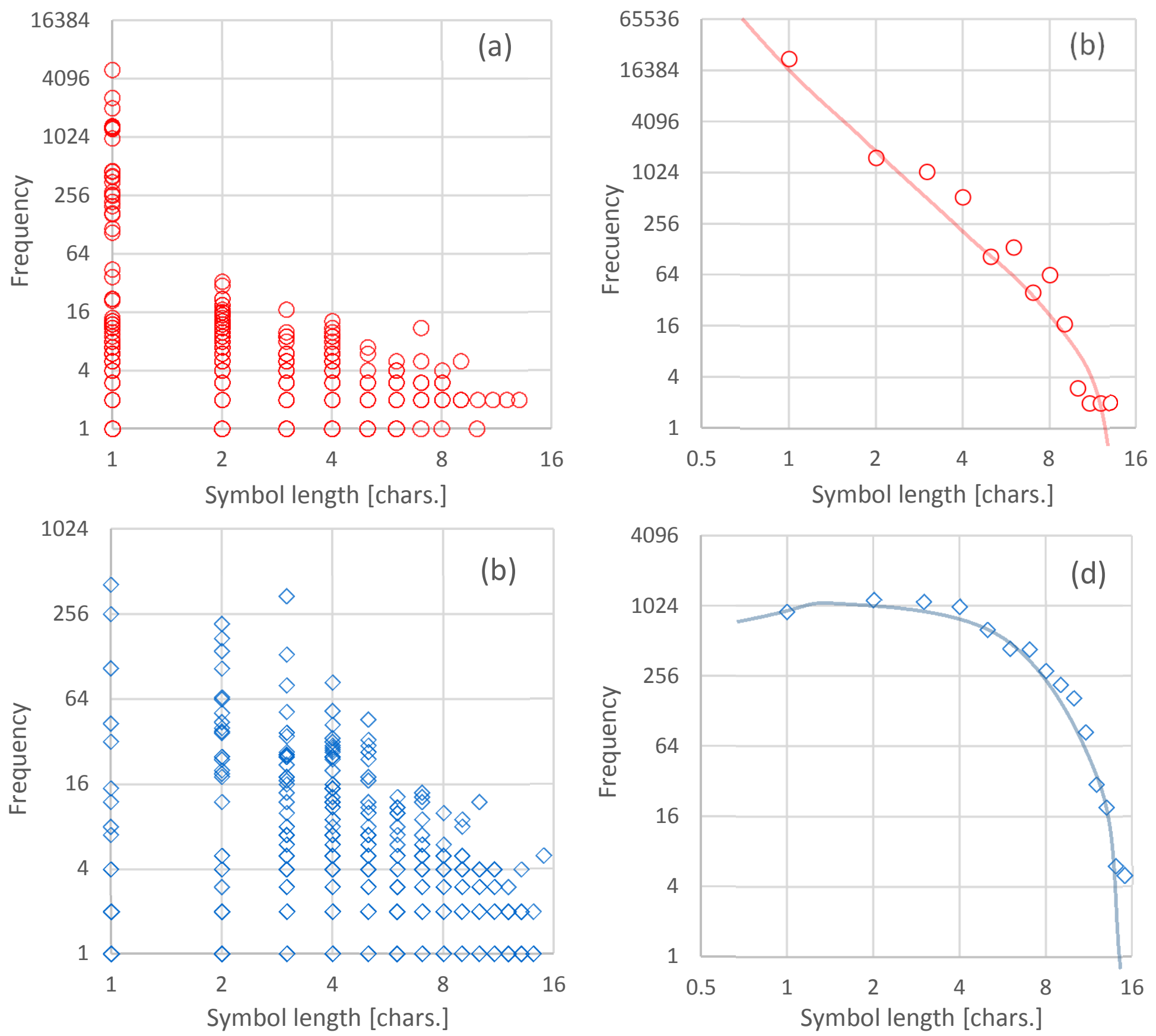

**Figure 4.** Bertrand Russell's 1950 Nobel ceremony speech behavior according symbol length. (**a**) At fundamental scale symbol occurrences vs symbol length. (**b**) At fundamental scale symbol-length frequency distribution. (**c**) At word-scale symbol occurrences *vs*. symbol length. (**d**) At word-scale symbol-length frequency distribution.

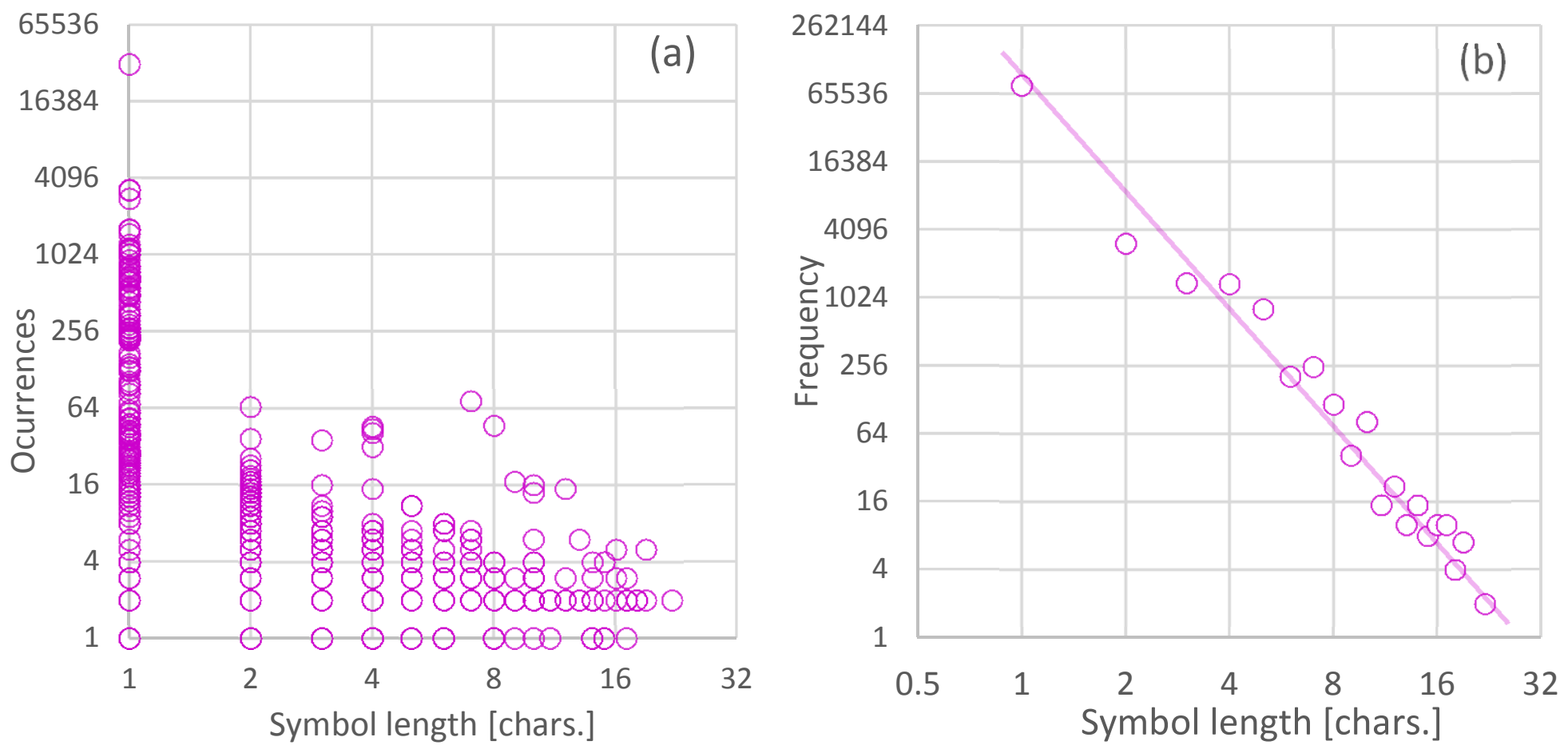

**Figure 5.** Beethoven's 9th symphony 4th movement MIDI music language behavior according symbol length. . (**a**) At fundamental scale symbol occurrences *vs*. symbol length. (**b**) At fundamental scale symbol-length frequency distribution.

Any scale of observation has a correspondence with the size of the symbols focused at that scale. When that size is the same for all symbols, the scale can be regarded as a regular scale and specified indicating its size. If on the contrary, the scale does not correspond to a constant symbol size, then a symbol frequency distribution based on the sizes is a valid depiction of the scale. That is the case of the scales of words for English texts and the fundamental scale for our two examples. Figures 4 and 5 show those distributions and are useful to interpret the fundamental scales of both examples.

## 5. Discussion

The results clearly showed the calculus of the entropy content of a communication system varies in important ways, depending on the scale of analysis. Looking at a language at the scale of characters provides a different picture than examining it at the level of words, or at the here described fundamental scale. Thus, in order to compare different communication systems, we need to use a similar scale applicable to each communication system. We showed that the fundamental scale presented here is applicable to very different communication systems, such as music, computer programs, and natural languages. This allows us to perform comparative studies regarding the systems entropy and thus to infer about the relative complexity of different communication systems.

In both examples analyzed, the profiles at the scale of characters and the fundamental scale run close to each other, within the range of the most frequent symbols to the symbols with a rank placed near the mid logarithmic scale. For points with lower ranking, the fundamental-scale profile extends its tail toward the region of low symbol frequencies. The closeness of fundamental and character scaled profiles in the high frequency region, indicates that the character-scaled language $\boldsymbol{B}_1$ is a subset of the fundamental scale language. The language at fundamental scale, having a greater symbolic diversity and therefore more degrees of freedom, finds a way to generate a symbol frequency distribution with a lower entropy as compared to the minimal entropy distribution when the description is viewed at the scale of words. Focusing in the fundamental scale profiles, the symbols located in the lower rank region—the tail of the profile—tend to be longer symbols formed by more than one character. These multi-character symbols, which cannot exist at the character scale, are formed at the expense of instances of single character symbols typically located in the profile's head. This explains the nearly constant gap between the two profiles in the profiles' heads.

The English description, observed at the scale of words, produces a symbol profile incapable of showing short symbols—fragments of a word—which would represent important aspects of a spoken language as syllabus and other typical fundamental language sounds. On the opposite extreme, by observing at the character scale, the profile forbids considering strings of characters as symbols, thus meaningful words or structures cannot appear at this scale, missing important information about the structure of the described system.

The fundamental scale, on the other hand, appears as an intermediate scale capable of capturing the essence of the most elementary structure of a language, as its alphabet, as well as larger structures which represent the result of language evolution in its way to form more specialized and complex symbols. The same applies for music MIDI representation. There is no word scale for music, but clearly the character scale does not capture the richness that undoubtedly is present in this type of language.

Another difference between the fundamental scale, and other scales is the sensitivity to the order of the symbols as they appear in the text. At the scale of words or the scale of characters, the symbol frequency profile does not vary as the symbol order. The profiles depend only on the number of appearances of each symbol, word or character, depending on the subject scale. The profile built at the fundamental scale does change as the symbol order is altered, not because of the symbol order itself, but because the symbol set recognized as fundamental, changes when the order or words or characters are modified. As a consequence, the character and word scales do not have any sense of grammar. The fundamental scale and its corresponding profile, on the other hand, is affected by the order in which words are organized—or disorganized—and is therefore sensitive to the rules of grammar. Other communication systems may not have words, but they must have some rules or the equivalence of a grammar. Assuming rigid rules as symbol size or symbol delimiters seems to be a barrier when studying the structure of system descriptions.

In the search for symbols, the fundamental scale method accounts for frequent sequences of strings which result from grammar rules. The string "*ing*", for example appears at the end of words representing verbs or actions. Moreover, it normally comes followed by a space character (" "). As the sequence appears with noticeable frequency, the fundamental scale method recognizes the char sequence "*ing*" (ending with a space) as an entropy reducer token and therefore an important descriptive piece of English as a language. The observation of a description at its fundamental scale is therefore, sensitive to the order in which char-strings appear within the description. The fundamental scale method detects the internal grammar which has been ignored when analyzing Zipf's profiles at the scale of words in many previous studies.

Despite the concept of fundamental scale being applicable to descriptions built over multidimensional spaces, the fundamental scale method and the algorithm developed is devised for 1-dimensional descriptions. The symbol search process implemented scans the description along the writing dimension of the text file being analyzed. This means that the fundamental symbols constituting 2D descriptions like pictures, photographs or plain data tables cannot be discovered with the algorithm as developed. To extend the fundamental scale algorithm to descriptions of more than one dimension, the restriction (8c) must be modified or complemented, to incorporate the sense of indivisible information unit—as has been the character in the development of this study—and the allowed symbol boundary shape in the description-space considered. This adjustment is a difficult task to accomplish because establishing criteria for the shapes of the boundaries becomes a hard to solve topology problem, especially in higher dimensional spaces.

There are other limitations for the analysis of descriptions of one dimension. Some punctuation signs which belong more to the writing system than to the language itself, work in pairs. Parenthesis, quotes, admiration and question marks are some of the written punctuation signs which work in couples. Intuition indicates that each one of them is a half-symbol belonging to one symbol. In these cases, not considering each half as part of the same symbol most likely increases the entropy associated to the set of symbols discovered, thus becoming a deviation of the ideal application of the method. Nevertheless, for English, Spanish and human natural languages, in general, the characters which work in couples, appear infrequently as compared to the rest of characters. Thus the minimal entropy distortion introduced by this effect is small.

Practical use of the algorithm is feasible up to some description lengths. The actual limit depends on the nature of the language used in the description. For syllabic human natural languages the algorithm can be directly applied to texts of 40,000 characters or less. Longer texts, however, can be analyzed by partitioning. Thus the application limit for texts expressed in human natural languages, covers most needs. For the analysis of music, the use of the algorithm is limited to the MIDI format, result in large processing times even for powerful computers available today. The problem of scanning all possible sets of symbols in a sequence of characters grows as a combinatorial number. The Problem rapidly gets too complex in the computational sense, and its practical application is only feasible for representations of music in reduced sets of digitized symbols like the MIDI coding. Using more comprehensive formats like .MP3, a compressing technology capable of reducing the size of a music pack while keeping reasonably good sound quality, would be enough to locate the solution of the problem beyond our possibilities of performing experiments with large sets of musical pieces. Yet, the fundamental scale method provides new possibilities for discovering the most representative dimension of small sized textual descriptions, allowing us to advance in our understanding of languages.

The Fundamental Scale, as a concept and as a method to find a quantitative approximation to the description of communication systems promises to be fruitful in further research. Tackling the barriers of the algorithm by finding ways to reduce the number of loops and augmenting the assertiveness of the criteria used, may extend the space of practical use of the notion of a description's fundamental scale. Here we showed that the method reveals structural properties of languages and other communication systems, offering a path for comparative studies of the complexity of communication.

**Acknowledgements**

We wish to thank to anonymous referees for comments which improved the presentation of our work.

**Author Contributions**

Conceived the notion of Fundamental Scale: Gerardo Febres. Conceived and implemented the algorithm: Gerardo Febres. Designed and performed the tests: Gerardo Febres. Analyzed the results: Gerardo Febres and Klaus Jaffe. Wrote the paper: Gerardo Febres and Klaus Jaffe. Both authors have read and approved the final manuscript.

## Appendix A. The Fundamental Scale Algorithm Pseudo-Code

The following are a series of pseudo-codes of routines to determine the Fundamental Scale of any sequence of characters.

> BlueLetterPhrases: refers to computer instructions: Routine names, control loops and conditional statements.
>
> ***BlackItalicLetterPhrases:*** refers to variables. 1D Arrays are followed by [], and 2D arrays are followed by [,].
>
> GreyLetterPhrases: Comments and NAME OF PROCESS STAGES.

```
FundamentalScale(TheText, MaximumSymbolSize, Symbol[], SymbolFrequency[], SymbolPosition[,])
Scans TheText looking for symbols formed by adjacent characters. Returns the array of different
symbols Symbol[], their frequency of appearance SymbolFrequency[] and their position of appearance
SymbolPosition[,] within TheText.

SymbolSize = 1
UncertLowerLimit = 1, UncertUpperLimit = 0
while SymbolSize ≤ MaximumSymbolSize
    " BASE LANGUAGE CONSTRUCTION (When SymbolSize = 1)
    ProcessTextForASymbolSize(TheText, SymbolSize, Symbol(I), SymbolFrequency[], SymbolPosition[,])

    " CONSTRUCTION OF LANGUAGES WITH LONGER SYMBOLS (When SymbolSize > 1)
    BirthAndSurvival(Symbol[], SymbolFrequency[], SymbolPosition[], PropectiveSymbol[],
        PropectSymbolFreq[], PropectSymbolPosition[], N)

    SymbolSize = SymbolSize + 1
end while
```

```
ProcessTextForASymbolSize( TheText, SymbolSize, Symbol[], SymbolFrequency[], SymbolPosition[,])
Scans TheText looking for a characters sequences of SymbolSize characters, varying the position of the
cursor at the beggining of the reading process. Returns the array of different symbols with size =
SymbolSiz , their frequency of appearance SymbolFrequency[] , and the SymbolPosition[,] at any scan

Phase = 0
while Phase ≤ SymbolSize
    ScanTextStartingAtAPhase(TheText, SymbolSize, Phase, Symbol[], SymbolFrequency[], SymbolPosition[,])
    Phase = Phase + 1
end while
for each i                                                   " for each ProspectiveSymbol[]
    N = N + SymbolFrequency[]
end for
ConsolidateSymbolsFromDifferentPhases(PropectiveSymbol[], PropectSymbolFreq[], PropectSymbolPosition[])
```

```
EntropyOfASymbolSet(SymbolFrequency[], Entropy)
Computes the entropy of the symbols present in set. Each symbol is present in the set with the quantity indicated in
the array SymbolFrequency[] .

Entropy = 0
N = UpperBound of array SymbolFrequency[] (+1 deppending on the coding language)
For i = 1 to N
    Entropy = Entropy - SymbolFrequency[i] / N · log (SymbolFrequency[i] / N )
endfor
Entropy = Entropy / log (N)
```

PROSPECTIVE SYMBOL DETECTION

**ScanTextStartingAtAPhase(** ***TheText, SymbolSize, Phase, Symbol[], SymbolFrequency[], SymbolPosition[,]*** **)**

Scans ***TheText*** looking for a chracaters sequences of ***SymbolSize*** characters, starting the reading at the character position ***Phase***. Returns the array of different ***Symbol[]*** with size = ***SymbolSize***, their frequency of appearance at this scan ***Phase***, ***SymbolFrequency[]***, and the array of ***SymbolPosition[,]***.

```
CursorPosition = Phase          " CursorPosition = the position of the cursor in the process of reading a text
StillSomeCharsToRead = true
if CursorPosition > TextLength - Phase then
    StillSomeCharsToRead = false
    i = 0
end if
while StillSomeCharsToRead
    SymbolJustRead = TheSequenceOfSymbolSize CharsReadAt CursorPosition
    ThisIsANewSymbol = true
    if SymbolJustRead IsNotAnElementOfArray Symbol[I] then ThisIsANewSymbol = false
    if ThisIsANewSymbol then
        i = i + 1
        Symbol(i) = SymbolJustRead
        SymbolFrequency[i] = 1
    else
        IndexOfExistingSymbol = IdentifyIndexOfSymbolJustRead
        SymbolFrequency[IndexOfExistingSymbol] = SymbolFrequency[IndexOfExistingSymbol] + 1
    end if
    CursorPosition = CursorPosition + Phase
    MoveCursorToPosition CursorPosition
    if CursorPosition > TextLength - Phase then StillSomeCharsToRead = false
end while
```

PROSPECTIVE SYMBOL OVERLAP REDUCTION

**ConsolidateSymbolsFromDifferentPhases(** ***PropectiveSymbol[], PropectSymbolFreq[], PropectSymbolPosition[,]***

Filters the intances of the ***PropectiveSymbols[]*** by deleting the instances being partially ovelapped by other, with higher priority ***PropectiveSymbols[]***.

```
OrderArrays ProspectiveSymbol[], ProspectSymbolFreq[], ProspectSymbolPosition[,] by the value of
    ProspectSymbolFreq[].                                " Higher Frequency is more prioritary.
i = 0
for each i                                               " for each ProspectiveSymbol[i]
    j = 0
    for each j                                           " for each PropectSymbolPosition[i,j]
        Locate the ProspectiveSymbol[k] and the instaces m affected by (conflicting with) the insertion
           of the PropectiveSymbol[i]                    " There may be more than one ProspectiveSymbol[] affected.
        for each k              " for each PropectiveSymbol[k] affected by insertion of ProspectiveSymbol[i].
            for each m          " for each instance m of the PropectiveSymbol[k] affected by some insertion.
                Delete the instace m of ProspectiveSymbol[k] located at ProspectSymbolPosition[k,m]
                Update arrays ProspectSymbolFreq[] and ProspectSymbolPosition[,]
            end for
        end for

    end for
end for
for each i                                               " for each ProspectiveSymbol[i]
    if ProspectSymbolFreq[] < 2
        Delete elements i of arrays ProspectiveSymbol[i], ProspectSymbolFreq[i] and ProspectSymbolPosition[i, ]
           and ProspectSymbolPosition[i, ]
    end if
end for
```

BIRTH AND SURVIVAL PROCESSES

**BirthAndSurvival(*Symbol[], SymbolFrequency[], SymbolPosition[], Propective Symbol[], PropectSymbolFreq[], PropectSymbolPosition[], N*)**

Inserts ***ProspectiveSymbol[i]*** into the arrays ***Symbol[]*** if favorable condirions for an entropy reduction are observed. Every time a ***ProspectiveSymbol[i]*** is inserted into the ***Symbol[]*** array, an entropy test is performed. If no entropy decrease is observed, the lastly inserted symbol is deleted and arrays are reverted to their condition prior to the insertion. Returns the updated arrays ***Symbol[], SymbolFrequency[], SymbolPosition[]*** .

```
i = 0
Entropy  = 0
EntropyOfASymbolSet(SymbolFrequency[], Entropy )
D = UpperBound of array Symbol[]  (+1 deppending on the coding language)
UncertaintyPerSymbol = Entropy / D
Lambda = 0.01
for each i                                                    " for each ProspectiveSymbol[]
    P[i] = SymbolFrequency[] / N
    PropectSymbolUncertainty[ i ]  = - P[i] * log(P[i]) / log(D)
    if UncertaintyPerSymbol - Lambda < PropectSymbolUncertainty[i]  < UncertaintyPerSymbol + Lambda
        BIRTH PROCESS
        j = 0
        for each PropectSymbolPosition[i,j]                   " for each PropectSymbolPosition[i,j]
            Copy established arrays
            EstablishedSymbol[] = Symbol[]
            EstablishedSymbolFrequency[] = SymbolFrequency[]
            EstablishedSymbolPosition[,] = SymbolPosition[,]

            CONSERVATION OF SYMBOLIC QUANTITY
            Locate the Symbol[k] and the instaces m affected by (conflicting with) the insertion
                of the Symbol[i]                              " There may be more than one Symbol[] affected
            for each k                    " for each Symbol[k] affected by insertion of ProspectiveSymbol[i]
                for each m                " for each instance m of Symbol[k] affected by some insertion.
                    Delete the instace m of Symbol[k] located at ProspectSymbolPosition[k,m]
                    Update arrays SymbolFrequency[] and SymbolPosition[,]
                end for
            end for
            Insert ProspectiveSymbol[i] into array Symbol[]
            Update arrays SymbolFrequency[] and SymbolPosition[,]

            SURVIVAL PROCESS
            EntropyOfASymbolSet(SymbolFrequency[], Entropy )
            if Entropy < EstablishedEntropy  then                      Entropy decreased
                EstablishedEntropy  = Entropy
            else                                                       Entropy increased
                          Reject ProspectiveSymbol just inserted and Revert to Previous arrays
                Symbol[] = EstablishedSymbol[]
                SymbolFrequency[] = EstablishedSymbolFrequency[]
                SymbolPosition[,] = EstablishedSymbolPosition[,]
            end if
        end for
        EntropyOfASymbolSet(SymbolFrequency[], Entropy )
        EstablishedEntropy  = Entropy
    else                                              " PropectSymbolUncertainty[i] out of band
                                                      " Prospective Symbol has no oportunity to survive
    end if
end for
```

## Appendix B

*B.1. Bertrand Russell's speech given at the 1950 Nobel Award Ceremony:*

Word-scale profile Complete List. Speech text.
Total number of symbols [words]: 5716. Diversity: 1868.

| Rank | Symbol | Occurrences | Length | Rank | Symbol | Occurrences | Length |
|---|---|---|---|---|---|---|---|
| 1 | , | 412 | 1 | 51 | or | 20 | 2 |
| 2 | the | 342 | 3 | 52 | some | 20 | 4 |
| 3 | . | 256 | 1 | 53 | no | 19 | 2 |
| 4 | of | 218 | 2 | 54 | so | 18 | 2 |
| 5 | to | 172 | 2 | 55 | was | 18 | 3 |
| 6 | is | 140 | 2 | 56 | our | 18 | 3 |
| 7 | and | 132 | 3 | 57 | human | 18 | 5 |
| 8 | a | 106 | 1 | 58 | can | 17 | 3 |
| 9 | in | 105 | 2 | 59 | these | 17 | 5 |
| 10 | that | 84 | 4 | 60 | very | 16 | 4 |
| 11 | are | 80 | 3 | 61 | may | 16 | 3 |
| 12 | it | 66 | 2 | 62 | many | 16 | 4 |
| 13 | be | 64 | 2 | 63 | ; | 15 | 1 |
| 14 | they | 53 | 4 | 64 | than | 15 | 4 |
| 15 | not | 52 | 3 | 65 | such | 15 | 4 |
| 16 | as | 51 | 2 | 66 | fear | 15 | 4 |
| 17 | which | 46 | 5 | 67 | motives | 14 | 7 |
| 18 | if | 44 | 2 | 68 | war | 14 | 3 |
| 19 | I | 43 | 1 | 69 | life | 13 | 4 |
| 20 | have | 42 | 4 | 70 | people | 13 | 6 |
| 21 | we | 40 | 2 | 71 | however | 13 | 7 |
| 22 | by | 38 | 2 | 72 | because | 12 | 7 |
| 23 | you | 37 | 3 | 73 | « | 12 | 1 |
| 24 | he | 37 | 2 | 74 | his | 12 | 3 |
| 25 | but | 37 | 3 | 75 | excitement | 12 | 10 |
| 26 | for | 35 | 3 | 76 | hate | 12 | 4 |
| 27 | will | 34 | 4 | 77 | most | 12 | 4 |
| 28 | their | 33 | 5 | 78 | your | 12 | 4 |
| 29 | ' | 32 | 1 | 79 | great | 12 | 5 |
| 30 | with | 32 | 4 | 80 | an | 12 | 2 |
| 31 | from | 30 | 4 | 81 | think | 11 | 5 |
| 32 | power | 30 | 5 | 82 | become | 11 | 6 |
| 33 | this | 29 | 4 | 83 | been | 11 | 4 |
| 34 | when | 28 | 4 | 84 | motive | 11 | 6 |
| 35 | would | 27 | 5 | 85 | herd | 11 | 4 |
| 36 | more | 27 | 4 | 86 | much | 11 | 4 |
| 37 | one | 27 | 3 | 87 | out | 10 | 3 |
| 38 | there | 27 | 5 | 88 | should | 10 | 6 |
| 39 | who | 26 | 3 | 89 | could | 10 | 5 |
| 40 | has | 26 | 3 | 90 | those | 10 | 5 |
| 41 | them | 25 | 4 | 91 | politics | 10 | 8 |
| 42 | men | 25 | 3 | 92 | vanity | 10 | 6 |
| 43 | do | 25 | 2 | 93 | political | 9 | 9 |
| 44 | at | 25 | 2 | 94 | were | 9 | 4 |
| 45 | all | 25 | 3 | 95 | upon | 9 | 4 |
| 46 | what | 25 | 4 | 96 | desires | 9 | 7 |
| 47 | on | 24 | 2 | 97 | wish | 8 | 4 |
| 48 | other | 24 | 5 | 98 | ? | 8 | 1 |
| 49 | love | 24 | 4 | 99 | man | 8 | 3 |
| 50 | had | 22 | 3 | 100 | desire | 8 | 6 |

| Rank | Symbol | Occurrences | Length | Rank | Symbol | Occurrences | Length |
|---|---|---|---|---|---|---|---|
| 201 | boredom | 4 | 7 | 301 | preference | 3 | 10 |
| 202 | time | 4 | 4 | 302 | various | 3 | 7 |
| 203 | better | 4 | 6 | 303 | type | 3 | 4 |
| 204 | while | 4 | 5 | 304 | obvious | 3 | 7 |
| 205 | gambling | 4 | 8 | 305 | sometimes | 3 | 9 |
| 206 | serious | 4 | 7 | 306 | sank | 3 | 4 |
| 207 | long | 4 | 4 | 307 | away | 3 | 4 |
| 208 | found | 4 | 5 | 308 | cause | 3 | 5 |
| 209 | hand | 4 | 4 | 309 | end | 3 | 3 |
| 210 | old | 4 | 3 | 310 | killed | 3 | 6 |
| 211 | taken | 4 | 5 | 311 | innocent | 3 | 8 |
| 212 | members | 4 | 7 | 312 | believe | 3 | 7 |
| 213 | destructive | 4 | 11 | 313 | themselves | 3 | 10 |
| 214 | above | 4 | 5 | 314 | desired | 3 | 7 |
| 215 | within | 4 | 6 | 315 | step | 3 | 4 |
| 216 | enemies | 4 | 7 | 316 | wars | 3 | 4 |
| 217 | French | 4 | 6 | 317 | kind | 3 | 4 |
| 218 | way | 4 | 3 | 318 | where | 3 | 5 |
| 219 | communists | 4 | 10 | 319 | passions | 3 | 8 |
| 220 | effective | 4 | 9 | 320 | instinctive | 3 | 11 |
| 221 | sympathy | 4 | 8 | 321 | brothers | 3 | 8 |
| 222 | self | 4 | 4 | 322 | feeling | 3 | 7 |
| 223 | Nation | 4 | 6 | 323 | Russians | 3 | 8 |
| 224 | selfishness | 4 | 11 | 324 | enemy | 3 | 5 |
| 225 | moralists | 4 | 9 | 325 | ways | 3 | 4 |
| 226 | general | 4 | 7 | 326 | conflict | 3 | 8 |
| 227 | although | 4 | 8 | 327 | altruistic | 3 | 10 |
| 228 | politicians | 4 | 11 | 328 | against | 3 | 7 |
| 229 | since | 4 | 5 | 329 | operation | 3 | 9 |
| 230 | ideologies | 4 | 10 | 330 | fall | 3 | 4 |
| 231 | government | 4 | 10 | 331 | hunger | 3 | 6 |
| 232 | account | 3 | 7 | 332 | history | 3 | 7 |
| 233 | population | 3 | 10 | 333 | rivalry | 3 | 7 |
| 234 | South | 3 | 5 | 334 | current | 2 | 7 |
| 235 | North | 3 | 5 | 335 | theory | 2 | 6 |
| 236 | books | 3 | 5 | 336 | psychology | 2 | 10 |
| 237 | sort | 3 | 4 | 337 | facts | 2 | 5 |
| 238 | person | 3 | 6 | 338 | constitutional | 2 | 14 |
| 239 | between | 3 | 7 | 339 | began | 2 | 5 |
| 240 | cannot | 3 | 6 | 340 | right | 2 | 5 |
| 241 | politician | 3 | 10 | 341 | average | 2 | 7 |
| 242 | frequently | 3 | 10 | 342 | income | 2 | 6 |
| 243 | causes | 3 | 6 | 343 | want | 2 | 4 |
| 244 | action | 3 | 6 | 344 | tell | 2 | 4 |
| 245 | another | 3 | 7 | 345 | heard | 2 | 5 |
| 246 | far | 3 | 3 | 346 | questions | 2 | 9 |
| 247 | too | 3 | 3 | 347 | remote | 2 | 6 |
| 248 | wholly | 3 | 6 | 348 | scientific | 2 | 10 |
| 249 | duty | 3 | 4 | 349 | constantly | 2 | 10 |
| 250 | sense | 3 | 5 | 350 | thinking | 2 | 8 |

| Rank | Symbol | Occurrences | Lengt h | Rank | Symbol | Occurrences | Length |
|---|---|---|---|---|---|---|---|
| 501 | designed | 2 | 8 | 551 | century | 2 | 7 |
| 502 | deceive | 2 | 7 | 552 | feel | 2 | 4 |
| 503 | condemn | 2 | 7 | 553 | hatred | 2 | 6 |
| 504 | form | 2 | 4 | 554 | strange | 2 | 7 |
| 505 | appropriate | 2 | 11 | 555 | methods | 2 | 7 |
| 506 | feared | 2 | 6 | 556 | best | 2 | 4 |
| 507 | fellow | 2 | 6 | 557 | thoroughly | 2 | 10 |
| 508 | leads | 2 | 5 | 558 | produced | 2 | 8 |
| 509 | exciting | 2 | 8 | 559 | ill | 2 | 3 |
| 510 | provide | 2 | 7 | 560 | treated | 2 | 7 |
| 511 | rabbits | 2 | 7 | 561 | Western | 2 | 7 |
| 512 | impulse | 2 | 7 | 562 | countries | 2 | 9 |
| 513 | big | 2 | 3 | 563 | sum | 2 | 3 |
| 514 | contain | 2 | 7 | 564 | expensive | 2 | 9 |
| 515 | small | 2 | 5 | 565 | Germans | 2 | 7 |
| 516 | enmity | 2 | 6 | 566 | victors | 2 | 7 |
| 517 | actual | 2 | 6 | 567 | secured | 2 | 7 |
| 518 | member | 2 | 6 | 568 | advantages | 2 | 10 |
| 519 | mechanism | 2 | 9 | 569 | B | 2 | 1 |
| 520 | nations | 2 | 7 | 570 | large | 2 | 5 |
| 521 | regards | 2 | 7 | 571 | disguised | 2 | 9 |
| 522 | international | 2 | 13 | 572 | conclusion | 2 | 10 |
| 523 | discovered | 2 | 10 | 573 | intelligence | 2 | 12 |
| 524 | degree | 2 | 6 | 574 | ladies | 2 | 6 |
| 525 | says | 2 | 4 | 575 | economic | 2 | 8 |
| 526 | am | 2 | 2 | 576 | nor | 2 | 3 |
| 527 | line | 2 | 4 | 577 | mankind | 2 | 7 |
| 528 | Rhine | 2 | 5 | 578 | court | 2 | 5 |
| 529 | essential | 2 | 9 | 579 | civilized | 2 | 9 |
| 530 | danger | 2 | 6 | 580 | dance | 2 | 5 |
| 531 | TRUE | 2 | 4 | 581 | none | 2 | 4 |
| 532 | might | 2 | 5 | 582 | killing | 2 | 7 |
| 533 | regard | 2 | 6 | 583 | Royal | 1 | 5 |
| 534 | Mother | 2 | 6 | 584 | Highness | 1 | 8 |
| 535 | Nature | 2 | 6 | 585 | Gentlemen | 1 | 9 |
| 536 | cooperation | 2 | 11 | 586 | chosen | 1 | 6 |
| 537 | easily | 2 | 6 | 587 | subject | 1 | 7 |
| 538 | schools | 2 | 7 | 588 | lecture | 1 | 7 |
| 539 | turning | 2 | 7 | 589 | tonight | 1 | 7 |
| 540 | cruelty | 2 | 7 | 590 | discussions | 1 | 11 |
| 541 | everyday | 2 | 8 | 591 | insufficient | 1 | 12 |
| 542 | atom | 2 | 4 | 592 | statistics | 1 | 10 |
| 543 | bomb | 2 | 4 | 593 | organization | 1 | 12 |
| 544 | wicked | 2 | 6 | 594 | set | 1 | 3 |
| 545 | rival | 2 | 5 | 595 | forth | 1 | 5 |
| 546 | hating | 2 | 6 | 596 | minutely | 1 | 8 |
| 547 | burglars | 2 | 8 | 597 | difficulty | 1 | 10 |
| 548 | disapprove | 2 | 10 | 598 | finding | 1 | 7 |
| 549 | attitude | 2 | 8 | 599 | able | 1 | 4 |
| 550 | irreligious | 2 | 11 | 600 | ascertain | 1 | 9 |

*B.2.* *Bertrand Russell's speech given at the 1950 Nobel Award Ceremony:*
Fundamental-scale profile: Complete profile. Speech text.
Total number of symbols [Fundamental_Symbols]: 25,362. Diversity: 1247.

| Rank | Symbol | Probability | Occurrences | Length | Rank | Symbol | Probability | Occurrences | Length |
|---|---|---|---|---|---|---|---|---|---|
| 1 | ø | 0.192475 | 5020 | 1 | 51 | pr | 0.000513 | 13 | 2 |
| 2 | e | 0.099270 | 2589 | 1 | 52 | B | 0.000512 | 13 | 1 |
| 3 | t | 0.077074 | 2010 | 1 | 53 | fo | 0.000512 | 13 | 2 |
| 4 | n | 0.050588 | 1319 | 1 | 54 | .øTh | 0.000485 | 13 | 4 |
| 5 | a | 0.050490 | 1317 | 1 | 55 | ot | 0.000479 | 12 | 2 |
| 6 | o | 0.049389 | 1288 | 1 | 56 | st | 0.000477 | 12 | 2 |
| 7 | i | 0.049202 | 1283 | 1 | 57 | ly | 0.000475 | 12 | 2 |
| 8 | s | 0.047820 | 1247 | 1 | 58 | ı | 0.000475 | 12 | 1 |
| 9 | h | 0.047428 | 1237 | 1 | 59 | ur | 0.000474 | 12 | 2 |
| 10 | r | 0.037977 | 990 | 1 | 60 | ll | 0.000473 | 12 | 2 |
| 11 | d | 0.017566 | 458 | 1 | 61 | if | 0.000472 | 12 | 2 |
| 12 | l | 0.017226 | 449 | 1 | 62 | co | 0.000471 | 12 | 2 |
| 13 | f | 0.015580 | 406 | 1 | 63 | as | 0.000471 | 12 | 2 |
| 14 | c | 0.015178 | 396 | 1 | 64 | S | 0.000471 | 12 | 1 |
| 15 | w | 0.013417 | 350 | 1 | 65 | E | 0.000469 | 12 | 1 |
| 16 | m | 0.010668 | 278 | 1 | 66 | to | 0.000439 | 11 | 2 |
| 17 | y | 0.009954 | 260 | 1 | 67 | politic | 0.000439 | 11 | 7 |
| 18 | , | 0.009707 | 253 | 1 | 68 | F | 0.000436 | 11 | 1 |
| 19 | u | 0.008401 | 219 | 1 | 69 | ra | 0.000433 | 11 | 2 |
| 20 | p | 0.007610 | 198 | 1 | 70 | ca | 0.000433 | 11 | 2 |
| 21 | g | 0.006424 | 168 | 1 | 71 | øf | 0.000432 | 11 | 2 |
| 22 | v | 0.006262 | 163 | 1 | 72 | ce | 0.000430 | 11 | 2 |
| 23 | b | 0.004462 | 116 | 1 | 73 | K | 0.000428 | 11 | 1 |
| 24 | . | 0.004066 | 106 | 1 | 74 | will | 0.000425 | 11 | 4 |
| 25 | I | 0.001695 | 44 | 1 | 75 | øb | 0.000399 | 10 | 2 |
| 26 | k | 0.001421 | 37 | 1 | 76 | um | 0.000398 | 10 | 2 |
| 27 | nd | 0.001258 | 33 | 2 | 77 | em | 0.000397 | 10 | 2 |
| 28 | be | 0.001149 | 30 | 2 | 78 | M | 0.000395 | 10 | 1 |
| 29 | ma | 0.000831 | 22 | 2 | 79 | av | 0.000395 | 10 | 2 |
| 30 | of | 0.000829 | 22 | 2 | 80 | ev | 0.000395 | 10 | 2 |
| 31 | A | 0.000827 | 22 | 1 | 81 | su | 0.000394 | 10 | 2 |
| 32 | x | 0.000826 | 22 | 1 | 82 | ol | 0.000394 | 10 | 2 |
| 33 | T | 0.000787 | 21 | 1 | 83 | ver | 0.000393 | 10 | 3 |
| 34 | un | 0.000747 | 19 | 2 | 84 | se | 0.000393 | 10 | 2 |
| 35 | us | 0.000713 | 19 | 2 | 85 | whic | 0.000390 | 10 | 4 |
| 36 | .øI | 0.000668 | 17 | 3 | 86 | woul | 0.000363 | 9 | 4 |
| 37 | by | 0.000633 | 17 | 2 | 87 | pp | 0.000357 | 9 | 2 |
| 38 | s, | 0.000629 | 16 | 2 | 88 | de | 0.000356 | 9 | 2 |
| 39 | mo | 0.000627 | 16 | 2 | 89 | im | 0.000356 | 9 | 2 |
| 40 | me | 0.000626 | 16 | 2 | 90 | ua | 0.000355 | 9 | 2 |
| 41 | ed | 0.000599 | 16 | 2 | 91 | ac | 0.000355 | 9 | 2 |
| 42 | ad | 0.000592 | 15 | 2 | 92 | op | 0.000355 | 9 | 2 |
| 43 | lo | 0.000592 | 15 | 2 | 93 | wi | 0.000354 | 9 | 2 |
| 44 | ve | 0.000588 | 15 | 2 | 94 | from | 0.000354 | 9 | 4 |
| 45 | om | 0.000587 | 15 | 2 | 95 | com | 0.000354 | 9 | 3 |
| 46 | ød | 0.000586 | 15 | 2 | 96 | øp | 0.000353 | 9 | 2 |
| 47 | W | 0.000553 | 14 | 1 | 97 | no | 0.000353 | 9 | 2 |
| 48 | ri | 0.000551 | 14 | 2 | 98 | hi | 0.000353 | 9 | 2 |
| 49 | ag | 0.000514 | 13 | 2 | 99 | so | 0.000353 | 9 | 2 |
| 50 | ; | 0.000513 | 13 | 1 | 100 | ho | 0.000352 | 9 | 2 |

| Rank | Symbol | Probability | Occurrences | Length | Rank | Symbol | Probability | Occurrences | Length |
|---|---|---|---|---|---|---|---|---|---|
| 776 | øgr | $7.87\times10^{-5}$ | 2 | 3 | 1001 | nm | $4.00\times10^{-5}$ | 1 | 2 |
| 777 | lec | $7.87\times10^{-5}$ | 2 | 3 | 1002 | fl | $4.00\times10^{-5}$ | 1 | 2 |
| 778 | lki | $7.87\times10^{-5}$ | 2 | 3 | 1003 | rø | $4.00\times10^{-5}$ | 1 | 2 |
| 779 | ødea | $7.87\times10^{-5}$ | 2 | 4 | 1004 | du | $4.00\times10^{-5}$ | 1 | 2 |
| 780 | ?øAn | $7.87\times10^{-5}$ | 2 | 4 | 1005 | 8 | $4.00\times10^{-5}$ | 1 | 1 |
| 781 | rimi | $7.87\times10^{-5}$ | 2 | 4 | 1006 | J | $4.00\times10^{-5}$ | 1 | 1 |
| 782 | day, | $7.87\times10^{-5}$ | 2 | 4 | 1007 | øu | $4.00\times10^{-5}$ | 1 | 2 |
| 783 | joym | $7.87\times10^{-5}$ | 2 | 4 | 1008 | nu | $4.00\times10^{-5}$ | 1 | 2 |
| 784 | stsø | $7.87\times10^{-5}$ | 2 | 4 | 1009 | ob | $4.00\times10^{-5}$ | 1 | 2 |
| 785 | firs | $7.87\times10^{-5}$ | 2 | 4 | 1010 | cke | $4.00\times10^{-5}$ | 1 | 3 |
| 786 | forø | $7.87\times10^{-5}$ | 2 | 4 | 1011 | oul | $4.00\times10^{-5}$ | 1 | 3 |
| 787 | notø | $7.87\times10^{-5}$ | 2 | 4 | 1012 | ari | $4.00\times10^{-5}$ | 1 | 3 |
| 788 | mora | $7.87\times10^{-5}$ | 2 | 4 | 1013 | til | $4.00\times10^{-5}$ | 1 | 3 |
| 789 | eøol | $7.87\times10^{-5}$ | 2 | 4 | 1014 | le. | $4.00\times10^{-5}$ | 1 | 3 |
| 790 | hand | $7.87\times10^{-5}$ | 2 | 4 | 1015 | mun | $4.00\times10^{-5}$ | 1 | 3 |
| 791 | zedøm | $7.87\times10^{-5}$ | 2 | 5 | 1016 | war | $4.00\times10^{-5}$ | 1 | 3 |
| 792 | løref | $7.87\times10^{-5}$ | 2 | 5 | 1017 | tho | $4.00\times10^{-5}$ | 1 | 3 |
| 793 | uchøa | $7.87\times10^{-5}$ | 2 | 5 | 1018 | rth | $4.00\times10^{-5}$ | 1 | 3 |
| 794 | produc | $7.87\times10^{-5}$ | 2 | 6 | 1019 | døw | $4.00\times10^{-5}$ | 1 | 3 |
| 795 | inøcon | $7.87\times10^{-5}$ | 2 | 6 | 1020 | slav | $4.00\times10^{-5}$ | 1 | 4 |
| 796 | lømake¹up | $7.87\times10^{-5}$ | 2 | 9 | 1021 | øint | $4.00\times10^{-5}$ | 1 | 4 |
| 797 | heødevilø | $7.87\times10^{-5}$ | 2 | 9 | 1022 | aløs | $4.00\times10^{-5}$ | 1 | 4 |
| 798 | shouldøbe | $7.87\times10^{-5}$ | 2 | 9 | 1023 | nalø | $4.00\times10^{-5}$ | 1 | 4 |
| 799 | y¹fiveømil | $7.87\times10^{-5}$ | 2 | 10 | 1024 | ty.ø | $4.00\times10^{-5}$ | 1 | 4 |
| 800 | seriousness | $7.87\times10^{-5}$ | 2 | 11 | 1025 | tern | $4.00\times10^{-5}$ | 1 | 4 |
| 801 | fromøboredom | $7.87\times10^{-5}$ | 2 | 12 | 1026 | erty | $4.00\times10^{-5}$ | 1 | 4 |
| 802 | not,øperhaps, | $7.87\times10^{-5}$ | 2 | 13 | 1027 | lyøf | $4.00\times10^{-5}$ | 1 | 4 |
| 803 | uch | $7.86\times10^{-5}$ | 2 | 3 | 1028 | ent, | $4.00\times10^{-5}$ | 1 | 4 |
| 804 | ry | $7.86\times10^{-5}$ | 2 | 2 | 1029 | ryøg | $4.00\times10^{-5}$ | 1 | 4 |
| 805 | arø | $7.84\times10^{-5}$ | 2 | 3 | 1030 | eøk | $3.96\times10^{-5}$ | 1 | 3 |
| 806 | ' | $7.84\times10^{-5}$ | 2 | 1 | 1031 | pulse | $3.96\times10^{-5}$ | 1 | 5 |
| 807 | llø | $7.83\times10^{-5}$ | 2 | 3 | 1032 | øwould | $3.96\times10^{-5}$ | 1 | 6 |
| 808 | rad | $7.83\times10^{-5}$ | 2 | 3 | 1033 | unl | $3.96\times10^{-5}$ | 1 | 3 |
| 809 | dr | $7.83\times10^{-5}$ | 2 | 2 | 1034 | you, | $3.96\times10^{-5}$ | 1 | 4 |
| 810 | D | $7.83\times10^{-5}$ | 2 | 1 | 1035 | oføex | $3.96\times10^{-5}$ | 1 | 5 |
| 811 | øun | $7.79\times10^{-5}$ | 2 | 3 | 1036 | igh | $3.96\times10^{-5}$ | 1 | 3 |
| 812 | wil | $7.79\times10^{-5}$ | 2 | 3 | 1037 | rav | $3.96\times10^{-5}$ | 1 | 3 |
| 813 | about | $7.79\times10^{-5}$ | 2 | 5 | 1038 | øev | $3.96\times10^{-5}$ | 1 | 3 |
| 814 | y,øa | $7.79\times10^{-5}$ | 2 | 4 | 1039 | øsp | $3.96\times10^{-5}$ | 1 | 3 |
| 815 | Gr | $7.79\times10^{-5}$ | 2 | 2 | 1040 | eøt | $3.96\times10^{-5}$ | 1 | 3 |
| 816 | oe | $7.79\times10^{-5}$ | 2 | 2 | 1041 | xci | $3.96\times10^{-5}$ | 1 | 3 |
| 817 | lov | $7.79\times10^{-5}$ | 2 | 3 | 1042 | let | $3.96\times10^{-5}$ | 1 | 3 |
| 818 | øMa | $7.79\times10^{-5}$ | 2 | 3 | 1043 | ud | $3.96\times10^{-5}$ | 1 | 2 |
| 819 | iew | $7.79\times10^{-5}$ | 2 | 3 | 1044 | nn | $3.96\times10^{-5}$ | 1 | 2 |
| 820 | agr | $7.79\times10^{-5}$ | 2 | 3 | 1045 | hr | $3.96\times10^{-5}$ | 1 | 2 |
| 821 | ivi | $7.79\times10^{-5}$ | 2 | 3 | 1046 | ug | $3.96\times10^{-5}$ | 1 | 2 |
| 822 | ,ø« | $7.79\times10^{-5}$ | 2 | 3 | 1047 | gg | $3.96\times10^{-5}$ | 1 | 2 |
| 823 | .øO | $7.79\times10^{-5}$ | 2 | 3 | 1048 | ead | $3.96\times10^{-5}$ | 1 | 3 |
| 824 | esu | $7.79\times10^{-5}$ | 2 | 3 | 1049 | ,øe | $3.96\times10^{-5}$ | 1 | 3 |
| 825 | oøi | $7.79\times10^{-5}$ | 2 | 3 | 1050 | eed | $3.96\times10^{-5}$ | 1 | 3 |

*B.3. Beethoven 9th Symphony, 4th movement:*
Fundamental -scale profile: Complete profile. Complete text. Listen MIDI Version.
Total number of symbols [Fundamental_Symbols]: 84645. Diversity: 2824.

| Rank | Symbol | Probability | Occurrence | Length | Rank | Symbol | Probability | Occurrence | Length |
|---|---|---|---|---|---|---|---|---|---|
| 1 | ² | 0.38332 | 32446 | 1 | 51 | 1 | 0.00287 | 243 | 1 |
| 2 | x | 0.03896 | 3298 | 1 | 52 | ] | 0.00278 | 235 | 1 |
| 3 | Φ | 0.03870 | 3276 | 1 | 53 | 5 | 0.00273 | 231 | 1 |
| 4 | n | 0.03320 | 2810 | 1 | 54 | + | 0.00267 | 226 | 1 |
| 5 | @ | 0.01921 | 1626 | 1 | 55 | [ | 0.00265 | 224 | 1 |
| 6 | ³ | 0.01916 | 1622 | 1 | 56 | A | 0.00259 | 219 | 1 |
| 7 | d | 0.01769 | 1497 | 1 | 57 | 0 | 0.00261 | 221 | 1 |
| 8 | 9 | 0.01454 | 1231 | 1 | 58 | ! | 0.00205 | 173 | 1 |
| 9 | 2 | 0.01359 | 1151 | 1 | 59 | . | 0.00189 | 160 | 1 |
| 10 | ? | 0.01358 | 1149 | 1 | 60 | 8 | 0.00167 | 142 | 1 |
| 11 | - | 0.01321 | 1118 | 1 | 61 | W | 0.00160 | 135 | 1 |
| 12 | é | 0.01304 | 1104 | 1 | 62 | P | 0.00158 | 134 | 1 |
| 13 | J | 0.01221 | 1033 | 1 | 63 | - | 0.00150 | 127 | 1 |
| 14 | E | 0.01212 | 1026 | 1 | 64 | D | 0.00148 | 125 | 1 |
| 15 | B | 0.01108 | 938 | 1 | 65 | ——— | 0.00123 | 104 | 1 |
| 16 | L | 0.00997 | 844 | 1 | 66 | ¿ | 0.00117 | 99 | 1 |
| 17 | Q | 0.00979 | 828 | 1 | 67 | Y | 0.00115 | 97 | 1 |
| 18 | N | 0.00944 | 799 | 1 | 68 | 3 | 0.00107 | 91 | 1 |
| 19 | / | 0.00899 | 761 | 1 | 69 | % | 0.00100 | 84 | 1 |
| 20 | 6 | 0.00877 | 742 | 1 | 70 | ——— | 0.00087 | 73 | 7 |
| 21 | ; | 0.00826 | 699 | 1 | 71 | # | 0.00081 | 69 | 1 |
| 22 | C | 0.00815 | 690 | 1 | 72 | °@ | 0.00078 | 66 | 2 |
| 23 | = | 0.00801 | 678 | 1 | 73 | w | 0.00073 | 62 | 1 |
| 24 | O | 0.00773 | 654 | 1 | 74 | ½ | 0.00070 | 59 | 1 |
| 25 | V | 0.00748 | 633 | 1 | 75 | $ | 0.00064 | 54 | 1 |
| 26 | K | 0.00746 | 631 | 1 | 76 | | 0.00064 | 54 | 1 |
| 27 | ã | 0.00671 | 568 | 1 | 77 | | 0.00063 | 53 | 1 |
| 28 | ’ | 0.00658 | 557 | 1 | 78 | | 0.00057 | 48 | 1 |
| 29 | 4 | 0.00635 | 537 | 1 | 79 | | 0.00057 | 48 | 1 |
| 30 | ¾ | 0.00600 | 508 | 1 | 80 | ——— | 0.00056 | 47 | 8 |
| 31 | Z | 0.00594 | 503 | 1 | 81 | ú | 0.00054 | 46 | 1 |
| 32 | 7 | 0.00590 | 499 | 1 | 82 | ——— | 0.00054 | 46 | 4 |
| 33 | G | 0.00582 | 492 | 1 | 83 | , | 0.00054 | 46 | 1 |
| 34 | I | 0.00576 | 488 | 1 | 84 | ——— | 0.00053 | 44 | 4 |
| 35 | ° | 0.00517 | 438 | 1 | 85 | ¤ | 0.00050 | 43 | 1 |
| 36 | & | 0.00454 | 385 | 1 | 86 | | 0.00050 | 43 | 1 |
| 37 | F | 0.00426 | 360 | 1 | 87 | 4¾'4 | 0.00049 | 41 | 4 |
| 38 | R | 0.00410 | 347 | 1 | 88 | *f* | 0.00048 | 41 | 1 |
| 39 | X | 0.00409 | 346 | 1 | 89 | | 0.00047 | 40 | 0 |
| 40 | | 0.00401 | 340 | 1 | 90 | _ | 0.00047 | 40 | 1 |
| 41 | * | 0.00369 | 312 | 1 | 91 | " | 0.00047 | 40 | 1 |
| 42 | | 0.00345 | 292 | 1 | 92 | let | 0.00046 | 39 | 1 |
| 43 | S | 0.00323 | 274 | 1 | 93 | ` | 0.00044 | 37 | 1 |
| 44 | T | 0.00322 | 273 | 1 | 94 | nn | 0.00044 | 37 | 1 |
| 45 | : | 0.00320 | 271 | 1 | 95 | @³ | 0.00043 | 37 | 2 |
| 46 | H | 0.00318 | 269 | 1 | 96 | °@³ | 0.00043 | 36 | 3 |
| 47 | f | 0.00302 | 255 | 1 | 97 | Á | 0.00042 | 36 | 1 |
| 48 | M | 0.00299 | 253 | 1 | 98 | í | 0.00042 | 36 | 1 |
| 49 | U | 0.00285 | 241 | 1 | 99 | -¾'- | 0.00038 | 32 | 4 |
| 50 | | 0.00285 | 241 | 1 | 100 | v | 0.00038 | 32 | 1 |

| Rank | Symbol | Probability | Occurrences | Length | Rank | Symbol | Probability | Occurrences | Length |
|---|---|---|---|---|---|---|---|---|---|
| 401 | S?²G | 0.00006 | 5 | 4 | 651 | ¿²ΦR¾ | 0.00004 | 3 | 5 |
| 402 | ²9Zn | 0.00006 | 5 | 4 | 652 | H?nM² | 0.00004 | 3 | 5 |
| 403 | d | 0.00006 | 5 | 2 | 653 | ²Y¾²M¾n V | 0.00004 | 3 | 8 |
| 404 | ²é | 0.00006 | 5 | 2 | 654 | "²,r.¾²"¾, | 0.00004 | 3 | 10 |
| 405 | † | 0.00006 | 5 | 1 | 655 | V | 0.00004 | 3 | 2 |
| 406 | ², | 0.00006 | 5 | 2 | 656 | [ | 0.00004 | 3 | 2 |
| 407 | Φ8x | 0.00006 | 5 | 3 | 657 | 3 | 0.00004 | 3 | 2 |
| 408 | ãΦ | 0.00006 | 5 | 2 | 658 | N | 0.00004 | 3 | 2 |
| 409 | ΦB | 0.00005 | 5 | 2 | 659 | 1 | 0.00004 | 3 | 2 |
| 410 | Φ’ | 0.00005 | 5 | 2 | 660 | 7 | 0.00004 | 3 | 2 |
| 411 | ã | 0.00005 | 5 | 2 | 661 | ;x²7x | 0.00004 | 3 | 5 |
| 412 | X³ | 0.00005 | 5 | 2 | 662 | GxãS | 0.00004 | 3 | 5 |
| 413 | &³ | 0.00005 | 5 | 2 | 663 | ;²²7 | 0.00004 | 3 | 5 |
| 414 | #³ | 0.00005 | 5 | 2 | 664 | %²Φ2x | 0.00004 | 3 | 5 |
| 415 | ?²-?n | 0.00005 | 4 | 5 | 665 | &²ΦUx | 0.00004 | 3 | 5 |
| 416 | -; | 0.00005 | 4 | 2 | 666 | ¾²ΦEx | 0.00004 | 3 | 5 |
| 417 | 1³2= | 0.00005 | 4 | 4 | 667 | ²&xã | 0.00004 | 3 | 5 |
| 418 | L³²@ | 0.00005 | 4 | 4 | 668 | ²E²ΦS | 0.00004 | 3 | 5 |
| 419 | 8³2D | 0.00005 | 4 | 4 | 669 | ¾xãQ | 0.00004 | 3 | 5 |
| 420 | 6³2B | 0.00005 | 4 | 4 | 670 | ΦJx²é | 0.00004 | 3 | 5 |
| 421 | édnN | 0.00005 | 4 | 4 | 671 | ãfX²² | 0.00004 | 3 | 5 |
| 422 | =³nU² | 0.00005 | 4 | 5 | 672 | Φ]x²Q | 0.00004 | 3 | 5 |
| 423 | ;³nS² | 0.00005 | 4 | 5 | 673 | ²Φ4x | 0.00004 | 3 | 4 |
| 424 | L?²C? | 0.00005 | 4 | 5 | 674 | xãfZ² | 0.00004 | 3 | 5 |
| 425 | d-N²ΦLd-L²ΦJd-J | 0.00005 | 4 | 15 | 675 | +xãQ | 0.00004 | 3 | 5 |
| 426 | E? | 0.00005 | 4 | 2 | 676 | ΦZx²] | 0.00004 | 3 | 5 |
| 427 | &²Φ¾ | 0.00005 | 4 | 4 | 677 | 2²Φ@x | 0.00004 | 3 | 5 |
| 428 | NxnN | 0.00005 | 4 | 4 | 678 | é/ | 0.00004 | 3 | 2 |
| 429 | 2-²Φ9-29²Φ---R | 0.00005 | 4 | 14 | 679 | 62 | 0.00004 | 3 | 2 |
| 430 | Sx | 0.00005 | 4 | 2 | 680 | 72 | 0.00004 | 3 | 2 |
| 431 | éx | 0.00005 | 4 | 2 | 681 | ãfI | 0.00004 | 3 | 3 |
| 432 | 2Q | 0.00005 | 4 | 2 | 682 | 7d²+d | 0.00004 | 3 | 5 |
| 433 | 3² | 0.00005 | 4 | 2 | 683 | ;x²@x | 0.00004 | 3 | 5 |
| 434 | //ãf | 0.00005 | 4 | 4 | 684 | O2²L2 | 0.00004 | 3 | 5 |
| 435 | /²-/ãf9 | 0.00005 | 4 | 7 | 685 | Gx²;x²V | 0.00004 | 3 | 7 |
| 436 | Q²ΦQxnQ | 0.00005 | 4 | 8 | 686 | ;/²é/²9/ | 0.00004 | 3 | 8 |
| 437 | Q/²E/ãfQ | 0.00005 | 4 | 8 | 687 | Qx | 0.00004 | 3 | 2 |
| 438 | ]x²Bx²9xn] | 0.00005 | 4 | 10 | 688 | L | 0.00004 | 3 | 2 |
| 439 | 7xn7²Φ-xnE | 0.00005 | 4 | 10 | 689 | J | 0.00004 | 3 | 2 |
| 440 | xã | 0.00005 | 4 | 2 | 690 | Φ- | 0.00004 | 3 | 2 |
| 441 | Φ] | 0.00005 | 4 | 2 | 691 | Bd | 0.00004 | 3 | 2 |
| 442 | @ã | 0.00005 | 4 | 2 | 692 | Y²²M² | 0.00004 | 3 | 5 |
| 443 | ²ã.¾ã | 0.00005 | 4 | 6 | 693 | ²;xnK | 0.00004 | 3 | 5 |
| 444 | .¾ãf | 0.00005 | 4 | 4 | 694 | ²2xã | 0.00004 | 3 | 5 |
| 445 | K¾nT | 0.00005 | 4 | 4 | 695 | ²/xnW | 0.00004 | 3 | 5 |
| 446 | Kãf5 | 0.00005 | 4 | 4 | 696 | °@³Z | 0.00004 | 3 | 5 |
| 447 | VZ²JZ | 0.00005 | 4 | 5 | 697 | ΦZxnZ | 0.00004 | 3 | 5 |
| 448 | é?².?ãf | 0.00005 | 4 | 7 | 698 | /xãf]² | 0.00004 | 3 | 6 |
| 449 | ?²:?ãf: | 0.00005 | 4 | 7 | 699 | :xn | 0.00004 | 3 | 3 |
| 450 | .¾²"¾,úJ | 0.00005 | 4 | 8 | 700 | C²² | 0.00004 | 3 | 3 |

| Rank | Symbol | Probability | Occurrences | Length | Rank | Symbol | Probability | Occurrences | Length |
|---|---|---|---|---|---|---|---|---|---|
| 2101 | 2dZ | 0.00002 | 2 | 4 | 2771 | 4?-4 | 0.00001 | 1 | 4 |
| 2102 | V² | 0.00002 | 2 | 4 | 2772 | ²4?- | 0.00001 | 1 | 4 |
| 2103 | Xd²P | 0.00002 | 2 | 4 | 2773 | 9d²- | 0.00001 | 1 | 4 |
| 2104 | !²²[ | 0.00002 | 2 | 4 | 2774 | -²Φ/ | 0.00001 | 1 | 4 |
| 2105 | 2dn& | 0.00002 | 2 | 4 | 2775 | -4²Φ | 0.00001 | 1 | 4 |
| 2106 | 4V² | 0.00002 | 2 | 4 | 2776 | ² | 0.00001 | 1 | 2 |
| 2107 | E²] | 0.00002 | 2 | 4 | 2777 | K²7Kn | 0.00001 | 1 | 5 |
| 2108 | ²OKn | 0.00002 | 2 | 4 | 2778 | K²NK | 0.00001 | 1 | 4 |
| 2109 | Bd² | 0.00002 | 2 | 4 | 2779 | éK²7 | 0.00001 | 1 | 4 |
| 2110 | °@³ | 0.00002 | 2 | 4 | 2780 | xE | 0.00001 | 1 | 2 |
| 2111 | ²ΦGK | 0.00002 | 2 | 4 | 2781 | QK² | 0.00001 | 1 | 3 |
| 2112 | :d²N | 0.00002 | 2 | 4 | 2782 | Xd | 0.00001 | 1 | 2 |
| 2113 | !dnU | 0.00002 | 2 | 4 | 2783 | OK | 0.00001 | 1 | 2 |
| 2114 | Ld²* | 0.00002 | 2 | 4 | 2784 | ² | 0.00001 | 1 | 2 |
| 2115 | ¾²²4 | 0.00002 | 2 | 4 | 2785 | dx | 0.00001 | 1 | 2 |
| 2116 | EdSE | 0.00002 | 2 | 4 | 2786 | dI | 0.00001 | 1 | 2 |
| 2117 | Z²N²oN | 0.00002 | 2 | 8 | 2787 | Un | 0.00001 | 1 | 2 |
| 2118 | K²NK²EKã | 0.00002 | 2 | 8 | 2788 | ã* | 0.00001 | 1 | 2 |
| 2119 | ²LKãfC²² | 0.00002 | 2 | 8 | 2789 | Jd | 0.00001 | 1 | 2 |
| 2120 | ²4dxX²²P | 0.00002 | 2 | 8 | 2790 | ¼ | 0.00001 | 1 | 1 |
| 2121 | CK²LK²GK | 0.00002 | 2 | 8 | 2791 | Q | 0.00001 | 1 | 2 |
| 2122 | ²;K²7Kn@ | 0.00002 | 2 | 8 | 2792 | ²/ | 0.00001 | 1 | 2 |
| 2123 | X²ÁO²[ | 0.00002 | 2 | 8 | 2793 | dn[ | 0.00001 | 1 | 3 |
| 2124 | [²O²"-d | 0.00002 | 2 | 9 | 2794 | 2dã | 0.00001 | 1 | 3 |
| 2125 | Φ=Kn=²²7² | 0.00002 | 2 | 9 | 2795 | J- | 0.00001 | 1 | 2 |
| 2126 | d²]dÉ2²Y& | 0.00002 | 2 | 9 | 2796 | ú@² | 0.00001 | 1 | 3 |
| 2127 | ãf=²Φ@Kãf | 0.00002 | 2 | 9 | 2797 | ²¿³ | 0.00001 | 1 | 3 |
| 2128 | ]²jZd²Qd²] | 0.00002 | 2 | 10 | 2798 | F | 0.00001 | 1 | 2 |
| 2129 | &d²2dIQ²V | 0.00002 | 2 | 10 | 2799 | [° | 0.00001 | 1 | 2 |
| 2130 | -, | 0.00002 | 2 | 2 | 2800 | °@ | 0.00001 | 1 | 3 |
| 2131 | …F | 0.00002 | 2 | 2 | 2801 | 4²Φ | 0.00001 | 1 | 3 |
| 2132 | ²6 | 0.00002 | 2 | 2 | 2802 | ¿³ | 0.00001 | 1 | 2 |
| 2133 | 2³ | 0.00002 | 2 | 2 | 2803 | ΦT | 0.00001 | 1 | 2 |
| 2134 | E- | 0.00002 | 2 | 2 | 2804 | $- | 0.00001 | 1 | 2 |
| 2135 | ²A | 0.00002 | 2 | 2 | 2805 | -ƒ | 0.00001 | 1 | 2 |
| 2136 | ²¿ | 0.00002 | 2 | 2 | 2806 | ¡ | 0.00001 | 1 | 1 |
| 2137 | R | 0.00002 | 2 | 2 | 2807 | -,ú | 0.00001 | 1 | 3 |
| 2138 | ²L | 0.00002 | 2 | 2 | 2808 | ãé | 0.00001 | 1 | 3 |
| 2139 | ° | 0.00002 | 2 | 2 | 2809 | ¿³ã | 0.00001 | 1 | 3 |
| 2140 | ã@ | 0.00002 | 2 | 3 | 2810 | ²²6 | 0.00001 | 1 | 3 |
| 2141 | ²²’ | 0.00002 | 2 | 3 | 2811 | Φ@³ | 0.00001 | 1 | 3 |
| 2142 | ²J² | 0.00002 | 2 | 3 | 2812 | 0-²$ | 0.00001 | 1 | 4 |
| 2143 | ³²L | 0.00002 | 2 | 3 | 2813 | …Fé² | 0.00001 | 1 | 4 |
| 2144 | ,úé | 0.00002 | 2 | 3 | 2814 | ³Φ | 0.00001 | 1 | 2 |
| 2145 | VE² | 0.00002 | 2 | 3 | 2815 | š | 0.00001 | 1 | 1 |
| 2146 | -²@ | 0.00002 | 2 | 3 | 2816 | Ó | 0.00001 | 1 | 1 |
| 2147 | ƒV7 | 0.00002 | 2 | 3 | 2817 | ~ | 0.00001 | 1 | 1 |
| 2148 | Φ@- | 0.00002 | 2 | 3 | 2818 |  | 0.00001 | 1 | 1 |
| 2149 | X°@ | 0.00002 | 2 | 3 | 2819 | ³‡6 | 0.00001 | 1 | 3 |
| 2150 | ²ã¤ | 0.00002 | 2 | 3 | 2820 | ²7³ | 0.00001 | 1 | 3 |

## Conflicts of Interest

The authors declare no conflict of interest.